\documentclass[a4paper,12pt]{article}

\usepackage{amsmath}
\usepackage[style=numeric-comp,bibstyle=numeric,sorting=none,url=false,natbib=true,maxbibnames=99,firstinits=true]{biblatex}

\usepackage{hyperref}
\usepackage[a4paper,left=2.5cm,right=2.5cm,top=3cm,bottom=2.5cm]{geometry}
\usepackage{authblk}
\usepackage{amsmath,bm}
\usepackage{graphicx}
\usepackage{mathrsfs}
\usepackage{amssymb}
\usepackage{physics}
\usepackage{tensor}
\usepackage{xcolor}
\usepackage{tikz}
\usepackage{pgfplots}
\usetikzlibrary{spy}
\pgfplotsset{grid style={line width=.1pt,dashed,lightgray}}
\usepackage{caption}
\usepackage{subcaption}

\DeclareCaptionLabelFormat{subfigure}{#2)}
\usepackage{amsthm}
\usepackage{booktabs,dcolumn,caption}
\usepackage{multirow}
\usepackage[english]{babel}
\usepackage{soul}
\usepackage{mathtools}

\usepackage[title]{appendix}

\usepackage{xstring}

\newtheoremstyle{remark}
{}
{}
{\textnormal}
{}
{\bfseries}
{.}
{ }
{\thmname{#1}\thmnumber{ #2}}%

\theoremstyle{plain}
\newtheorem{remark}{Remark}

\title{On analytical integration of interaction potentials between cylindrical and rectangular bodies with a focus on van der Waals attraction}

\author[1,2]{A.~Borković}
\author[1]{M.~H.~Gfrerer}
\author[3,4,5]{R.~A.~Sauer}

\affil[1]{Institute of Applied Mechanics, Graz University of Technology, Technikerstraße 4/II, 8010 Graz, Austria, aleksandar.borkovic@aggf.unibl.org, aborkovic@tugraz.at}
\affil[2]{University of Banja Luka, Faculty of Architecture, Civil Engineering and Geodesy, Department of Mechanics and Theory of Structures, 78000 Banja Luka, Bosnia and Herzegovina}
\affil[3]{Institute for Structural Mechanics, Ruhr University Bochum, Universitätsstraße 150, 44801 Bochum }

\affil[4]{Department of Structural Mechanics, Gdansk University of Technology, ul. Narutowicza 11/12, 80-233 Gdansk, Poland} 

\affil[5]{Dept. of Mechanical Engineering, Indian Institute of Technology Guwahati, Assam 781039, India }
\date{}                     
\begin{document}
	
\newcommand{\ve}[1]{\boldsymbol{#1}} 
	
\maketitle
	
\section*{Abstract}

The paper deals with the analytical integration of interaction potentials between specific geometries such as disks, cylinders, rectangles, and rectangular prisms. Interaction potentials are modeled as inverse-power laws with respect to the point-pair distance, and the complete body-body potential is obtained by pairwise summation (integration). Several exact new interaction laws are obtained, such as disk-plate and (in-plane) rectangle-rectangle for an arbitrary exponent, and disk-disk and rectangle-rectangle for van der Waals attraction. To balance efficiency and accuracy, additional approximate laws are proposed for disk-disk, point-cylinder, and disk-cylinder interactions. A brief numerical example illustrates the application of the pre-integrated Lennard-Jones disk-disk interaction potential for the interaction between elastic fibers.

\textbf{Keywords}: interaction potential; van der Waals attraction; pairwise summation; disk-disk interaction; disk-cylinder interaction; disk-plate interaction; rectangle-rectangle interaction

\section{Introduction}
\label{secintro}


Intermolecular interactions are the underlying mechanism that governs a myriad of macro phenomena \cite{2011israelachvili}. The exact physics behind these interactions is complex and yet to be fully understood, but the electromagnetic potential is considered the fundamental cause. Intermolecular interactions are often modeled as an inverse-power law of the point-pair distance. When the exponent is greater than 3 and the bodies are in close proximity, the total interaction is dominated by the few closest-point pairs. The computational modeling of this short-range effect is both crucial and demanding, due to the competition of repulsive and attractive forces that define a contact. On the other hand, the modeling of long-range and transitional effects is less involved but can be equally important since these forces can bring the objects in proximity in the first place.


The most important interaction is van der Waals (vdW) attraction as it is responsible for practically all phenomena involving intermolecular interactions. 
The vdW forces act even between neutral molecules since they are caused by fluctuating charge distributions. Furthermore, they contribute to both small and large separation regimes. The accurate modeling of vdW interaction is quite involved since it consists of several contributions, depends on the retardation time, and is non-additive \cite{2005parsegian}. We are here mainly referring to the simplified case of non-retarded vdW interaction for which the pairwise summation approach is valid and that can be modeled with the point-pair inverse-power law with exponent 6.

The history of the pairwise summation (integration) of interaction potentials goes back to Hamaker, who derived a sphere-sphere vdW law in \cite{1937hamakera} and introduced what would be later known as the \emph{Hamaker constant}. The main assumption behind the point-pair summation approach is that a physical constant, valid for both point-pair and body-body interactions, exists. Almost two decades later, Lifshitz showed that this assumption does not hold \cite{2005parsegian}. However, by assuming small differences in material electromagnetic properties and neglecting the finite velocity of light, the so-called \emph{happy convergence} of the Hamaker and Lifshitz theories follows \cite{2005parsegian}. The main difference is that the Hamaker constant needs to be computed from bulk material properties but the geometric influence of the pairwise integration is still crucial. The pairwise summation of vdW attraction is readily used for a qualitative description of various phenomena, from bio-molecular to crystallization, while more appropriate models are still under consideration \cite{2015reilly}. 


The numerical modeling of intermolecular interactions is often based on molecular dynamics or Monte Carlo simulations. Molecular dynamics applies the laws of motion to each molecule which makes it computationally expensive, while Monte Carlo is a statistical method that cannot trace the actual movement of molecules \cite{2011israelachvili}. An alternative to these well-established approaches is the \emph{coarse-grained} model that is based on homogenization and coarse-graining of the molecular model. By utilizing the physics of molecular interactions with the efficacy of continuum contact formulations \cite{2006wriggers}, the coarse-grained model provides a good balance between accuracy and efficiency \cite{1997argento, 2007sauer, 2008sauer, 2016fan}. The idea of separating the interactions that occur within the body (intrasolid) and between the bodies (intersolid) allows us to represent the interaction potential between two bodies as a function of the gap vector.

The main motivation for the present research is to enable efficient and accurate numerical simulation of intermolecular interactions between slender deformable bodies using the coarse-grained approach. Since the numerical integration of two folded 3D integrals is computationally expensive, we are pursuing an analytical pre-integration to improve the computational efficiency and balance it with accuracy. Therefore, we focus on section-section interaction pairs, which can be utilized for beam-beam interactions \cite{2020grill,2021grill,2023meier,2024borkovićd}. However, other interaction pairs can be of interest, such as section-beam \cite{2023grill,2024grill}, section-plate, etc. We aim to find analytical expressions for the interaction potentials between specific geometrical objects that will allow efficient numerical simulations of interactions between deformable bodies. These laws were rarely studied before because the idea of using the coarse-grained model within structural beam and shell theories was conceived only recently \cite{2014sauer, 2020grill, 2024mokhalingam}. The present study focuses on pure geometrical considerations, while the actual values of physical constants, potentials, and forces are not considered. 




Let us briefly review available contributions regarding the interaction between bodies of interest. VdW attraction between rectangular prisms with various orientations is considered in \cite{1954vold}. It is emphasized that approximations are necessary to solve the six-folded integral. Several expressions for vdW and steric interactions between equally-sized rectangles and parallelopipeds are derived in \cite{1960derocco}. These laws are used in \cite{2023esquivel-sirvent} to model the finite-size effect of plate-plate interaction within the framework of a hybrid Hamaker-Lifshitz approach. Additionally, they are employed in \cite{2024casarella} to create a dataset for the modeling of interaction between clay platelets. 

A seminal work on vdW interaction between cylinders is given in \cite{1972langbein}, where the in-plane disk-disk law for an arbitrary exponent is derived in the form of a double infinite series. By focusing on vdW interaction, the same expression is derived in \cite{1973brenner}. An application of pairwise vdW summation to the surfaces of biological interest and its accuracy are discussed in \cite{1977nira}. Although the Hamaker approach overestimates the interaction potential, it gives the correct results if the orientation effects are considered.




An approximate point-cylinder vdW law is derived in \cite{2000montgomerya} to obtain a disk-cylinder interaction. The exact point-cylinder law is found in \cite{2003kirsch} and further integrated to obtain the interaction between a sphere and an infinite cylinder, considering both non-retarded an retarded vdW limits. The exact point-cylinder vdW law is also considered in \cite{2004oh} but the cylinder-cylinder interaction is found by numerical integration. An approach that can be considered as a generalization of the Derjaguin approximation is developed in \cite{2012dantchev} for the pairwise interaction between a 3D body and a half-space.

When an appropriate interaction potential law for all separations is not available, one approximate approach is to find asymptotic expressions for small and large separations, and then to interpolate these limits. This is done in \cite{2007rajter} for cylinder-substrate and cylinder-cylinder vdW interactions using the Lifshitz theory and the pairwise summation as a guide for the definition of an interpolating function.









\begin{figure}[h!]
	\begin{center}
		\includegraphics[width=\linewidth]{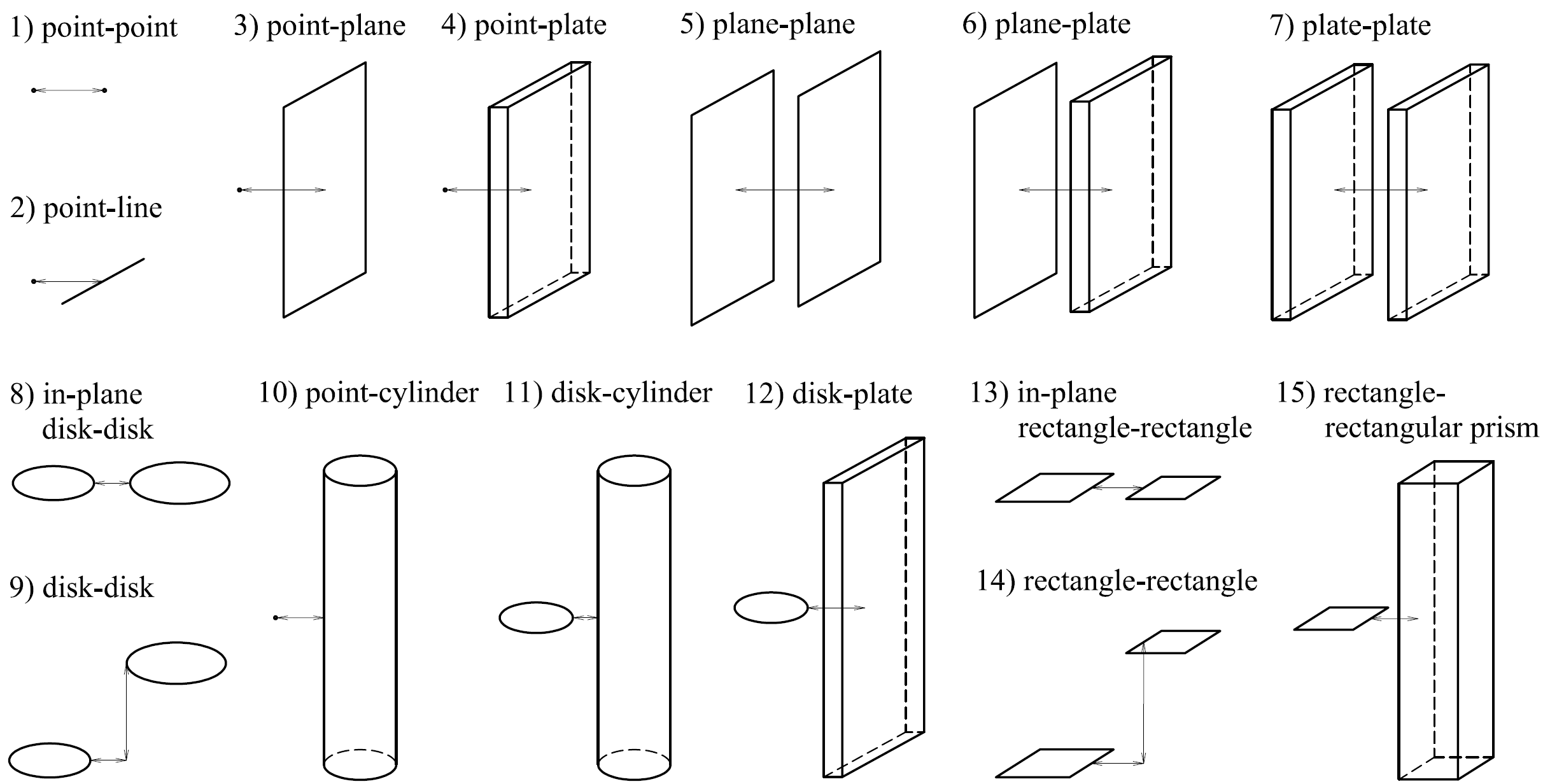}
		\caption{Various interaction pairs considered in this research. Due to parallel orientation and symmetry, the distance between all interaction bodies is defined by a gap (horizontal double arrow) and, for disk-disk and rectangle-rectangle, by an offset (vertical double arrow).}
		\label{fig:intro0}
	\end{center}
\end{figure}

Many of the existing expressions are summarized in \cite{2005parsegian,2000montgomerya,2011israelachvili}. In this research, we aim to derive new laws and to improve existing ones. 
The main contributions of the present paper are the derivation of exact vdW disk-disk and rectangle-rectangle laws, and an exact disk-plate law for a general exponent. Also, novel approximate point-cylinder and disk-cylinder laws are conceived. To derive these expressions, many different interaction pairs have to be considered and most of them are displayed in Fig.~\ref{fig:intro0}.
All the obtained expressions are verified with numerical integration and, where available, by comparison with the literature. As the main computational tool, we have used the computer algebra system Wolfram Mathematica 14 (WM14) \cite{2024wolframresearchinc.}. 
All the derivations are collected in six WM14 notebooks available as \nameref{supsec}.

To summarize, the subject of this paper is the modeling of intermolecular interactions between various bodies. Due to the underlying complexity of the problem, the following assumptions are introduced:
\begin{itemize}
	\item The point-point interaction potential is modeled as an inverse-power law of the point-pair distance.
	\item The total body-body interaction equals the pairwise summation (integration) of point-pair interactions.
	\item Only two-body interaction is considered and many-body effects are ignored.
	\item The interacting bodies are rigid, and in parallel and symmetric orientation, cf. Fig.~\ref{fig:intro0}. 
	\item Any influence of a surrounding medium is neglected.
	\item There is no redistribution of particles or charges inside the bodies; that is, we are dealing with dielectric or nonconducting materials. 
	\item The density distributions of particles and physical constants over the interacting bodies are constant.
\end{itemize}
Furthermore, let us state desired properties for exact interaction potential laws: (i) compact form, (ii) accurate evaluation with machine precision, (iii) real-valued and real arguments, (iv) general, i.e.~valid for an arbitrary exponent. For approximate laws, we introduce additional requirements: (v) good approximation for small separations, (vi) small and bounded errors for large separations, (vii) correct asymptotic scaling. Satisfaction of these requirements is difficult to achieve in general cases, but they can serve as general guidelines.

The remaining paper is organized as follows: a general concept of pairwise integration, coordinate systems and three rules are discussed and defined in the next section. Circular geometries are considered in the third and rectangular geometries in the fourth section. A compact numerical example is given in Section 5, which is followed by conclusions.

\section{General considerations}
\label{secgeneral}

In this section, we introduce the concept of the point-pair interaction and its integration over the two interacting bodies. Three rules and several general expressions are derived and revised as an introduction and foundation for the novel integrations that follow in Sections \ref{circsec} and \ref{rectsection}. Some commonly used abbreviations throughout the paper are: P for point, B for body, L for line, PN for plane, PT for plate, HS for half-space, R for rectangle, HST for half-strip, D for disk, C for cylinder, RP for rectangular prism, and S for sheet. Although some geometric objects are infinite by definition, such as plane and half-space, we will emphasize the infinity property in this section to be consistent when dealing with bodies that can have both infinite and finite dimensions, such as plate, cylinder, prism, etc.

Let us observe two bodies, $X$ and $Y$, that interact via some volume interaction potential. We exclusively consider point-pair interactions that can be represented as an inverse power law of $m^{th}$ order of the point-pair distance $r$, i.e.
\begin{equation}
		\Pi_{\operatorname{{P-P}}}^m = k_m \, r^{-m},
	\label{eq0}
\end{equation}
where $k_m$ is a physical constant, taken to be 1 in the following. An attractive potential is often defined as negative to obtain a positive force \cite{2011israelachvili}. We will consider all potentials as positive here, without loss of generality. To find a body-body interaction potential, the point-pair law must be integrated over both volumes, i.e.
\begin{equation}
	\begin{aligned}
		\Pi_{\operatorname{B-B}}^m&=\int_{V_x} \int_{V_y} \beta_x \, \beta_y \, \Pi_{\operatorname{{P-P}}}^m \dd{V_y} \dd{V_x},
	\end{aligned}
	\label{eq1rule0}
\end{equation}
%
%
%
where $\beta_i$ are the densities of particles in the respective bodies, taken to be 1 in the following. 

\begin{figure}[h!]
\begin{center}
	\includegraphics[width=0.6\linewidth]{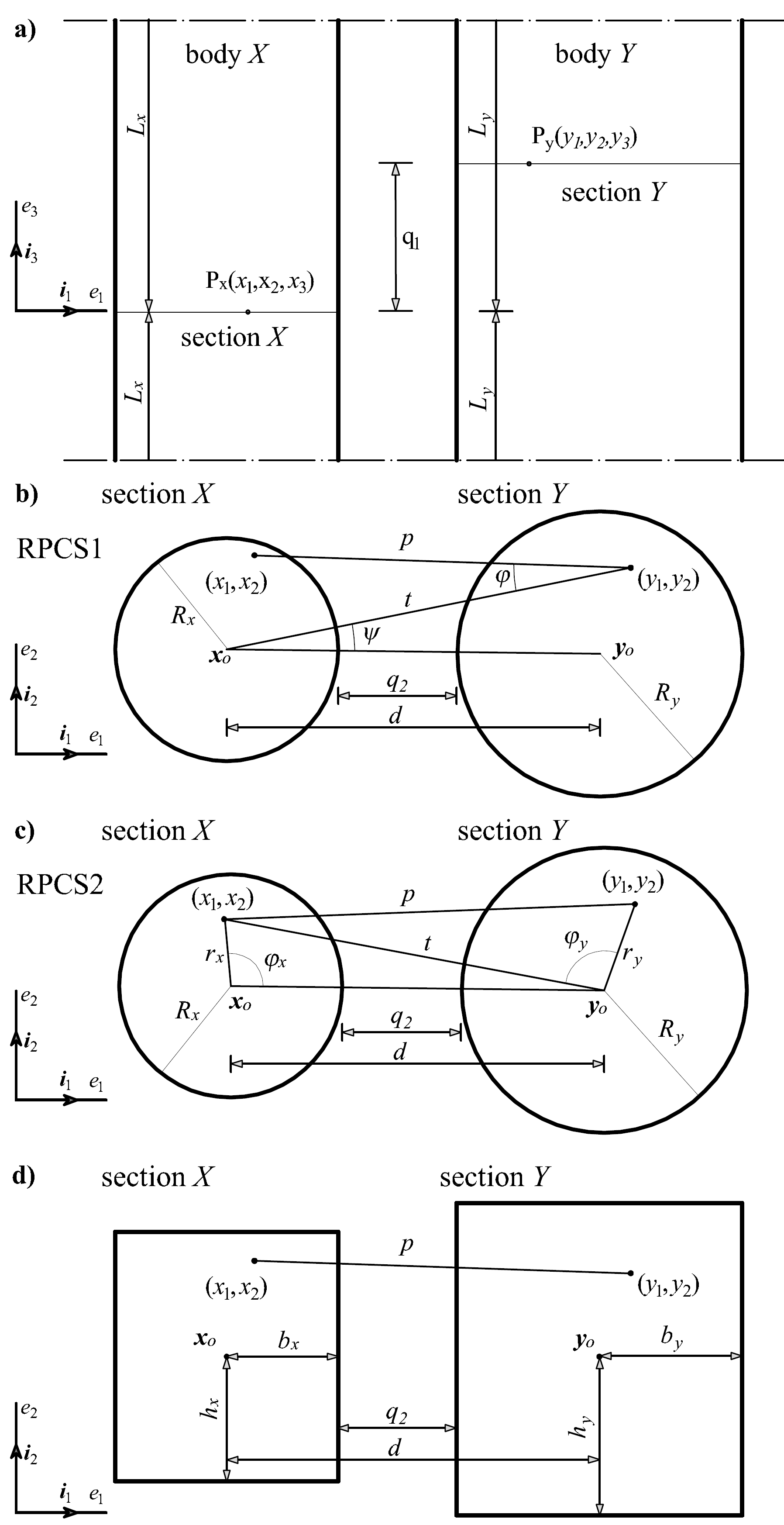}
	\caption{Interaction of two bodies with constant cross sections. a) Front view. b) and c) Top view in case of circular cross sections (disks). 
		Two relative polar coordinate systems are employed (RPCS1 and RPCS2). d) Top view in case of rectangular cross sections.
		}
		\label{fig:intro1}
	\end{center}
\end{figure}

The main subject of this research are interactions between parallel rigid cylinders, rectangular prisms, and their cross sections, see Fig.~\ref{fig:intro1}. 
For the integration over a circular domain we mainly use the relative polar coordinate system (RPCS1) of Fig.~\ref{fig:intro1}b \cite{1972langbein,2024borkovićd}, while for the rectangular domain, we employ the Cartesian coordinates of Fig.~\ref{fig:intro1}d. If not stated differently, the origin of the coordinate system is at the center of the body $X$, and denoted $\ve{x}_0$. Therefore, the coordinates of two arbitrary interacting points are $P_x=(x_1, x_2, 0)$ and $P_y=(y_1,y_2,q_1)$, since the offset between the cross sections is $q_1=y_3-x_3=y_3-0$. Then, the distance between the points is $r = \sqrt{p^2 + q_1^2}$, where $p$ is the in-plane distance between graphical projections of points into the $e_1e_2$ plane. For the Cartesian coordinate system, this distance is simply $p =\sqrt{(y_1 - x_1)^2 + (y_2 - x_2)^2}$, while for the polar coordinate systems in Fig.~\ref{fig:intro1}b and Fig.~\ref{fig:intro1}c, it is one of the coordinates.

Let us start with the integration along the $e_3$ axis.
Integrating the point-pair law along $q_1$ results in a $\operatorname{P-L_{\infty 3}}$ interaction, i.e. a point-infinite line (along the axis $e_3$) law, i.e. 
\begin{equation}
	\begin{aligned}
		\Pi_{\operatorname{P-L_{\infty 3}}}^m&=\int_{-\infty}^{\infty} \Pi_{\operatorname{{P-P}}}^m\dd{y_3}  =\int_{-\infty}^{\infty}(p^2+q_1^2)^{-\frac{m}{2}}  \dd{q_1} =\frac{\sqrt{\pi } \Gamma \left(\frac{m-1}{2}\right)}{\Gamma \left(\frac{m}{2}\right)}  \frac{1}{p^{m-1}} = f_m \Pi_{P-P_{IP}}^{m-1}, \\
		f_m & :=\frac{\sqrt{\pi } \Gamma \left(\frac{m-1}{2}\right)}{\Gamma \left(\frac{m}{2}\right)}, \quad m>1,
	\end{aligned}
\label{eq1rule}
\end{equation}
where $\Gamma\left(z\right)= \int_{0}^{\infty} p^{z-1} e^{-w} \dd{w}$ is the gamma function. This initial integration w.r.t.~the offset is used as a starting point in \cite{1972langbein} for the special case $m = 6$. 
The obtained relation \eqref{eq1rule} can be generalized in the following way: Observe an interaction between two in-plane objects, $X_{\text{IP}}$ and $Y_{\text{IP}}$, that lie in a common plane $\gamma$. These objects can be either points, plane lines, or plane figures/shapes (sections). Their interaction potential of $m^{th}$ order is $\Pi_{X_{\text{IP}}\text{-}Y_{\text{IP}}}^m$. Next, consider the interaction potential of $m^{th}$ order between the same body $X_{\text{IP}}$ and a new body $Y_{\infty}$ that is formed by infinitely extending the object $Y_{\text{IP}}$ perpendicular to the plane $\gamma$ in both directions, $\Pi_{X_{\text{IP}}\text{-}Y_{\infty}}^m$. For example, if $Y_{\text{IP}}$ is a point, $Y_{\infty}$ is an infinite line, if $Y_{\text{IP}}$ is a disk, $Y_{\infty}$ is an infinite cylinder, etc. The expression \eqref{eq1rule} implies that the interaction potential $\Pi_{X_{\text{IP}}\text{-}Y_{\infty}}^m$ can be expressed as a product of the in-plane potential of $(m-1)^{th}$ order, $\Pi_{X_{\text{IP}}\text{-}Y_{\text{IP}}}^{m-1}$, and a factor $f_m$, i.e.
\begin{equation}
	\begin{aligned}
		\Pi_{X_{\text{IP}}\text{-}Y_{\infty}}^m&=\frac{\sqrt{\pi } \Gamma \left(\frac{m-1}{2}\right)}{\Gamma \left(\frac{m}{2}\right)}  \Pi_{X_{\text{IP}}\text{-}Y_{\text{IP}}}^{m-1} = f_m \Pi_{X_{\text{IP}}\text{-}Y_{\text{IP}}}^{m-1}. 
	\end{aligned}
	\label{eq1rule1}
\end{equation}
We will refer to this observation as \emph{the rule of infinite and in-plane body interaction} (RIIPI). It will be readily used throughout the paper. 

\begin{remark}
	\normalfont
	The integration of interaction potentials between two bodies is often done by assuming that one of them is infinite along some dimension \cite{2011israelachvili}. In that case, the definite integral is much easier to evaluate and the result can be applied to many practical situations. For example, a point-finite line law for arbitrary $m$ involves a hypergeometric function, which makes it much more complicated for the evaluation and further integration than the point-infinite line law \eqref{eq1rule}, cf. \hyperref[supsec]{Notebook 1}.
\end{remark}

Let us find a point-infinite plane law. A straightforward approach is to apply RIIPI \eqref{eq1rule1} to the point-infinite line law \eqref{eq1rule}, i.e.
\begin{equation}
	\begin{aligned}
		\Pi_{\operatorname{P-PN_{\infty 23}}}^{m} = f_m \Pi_{\operatorname{P-L_{\infty 3}}}^{m-1} = \frac{2 \pi }{\left(m-2\right)\left(y_1-x_1\right)^{m-2}} = \frac{2 \pi }{\left(m-2\right)p^{m-2}}\; \text{for} \; m>2,
	\end{aligned}
	\label{eq1rule2}
\end{equation}
where the distance (gap) between the point and the plane is $p=y_1-x_1$. 

\begin{figure}[h!]
		\begin{center}
			\includegraphics[width=\linewidth]{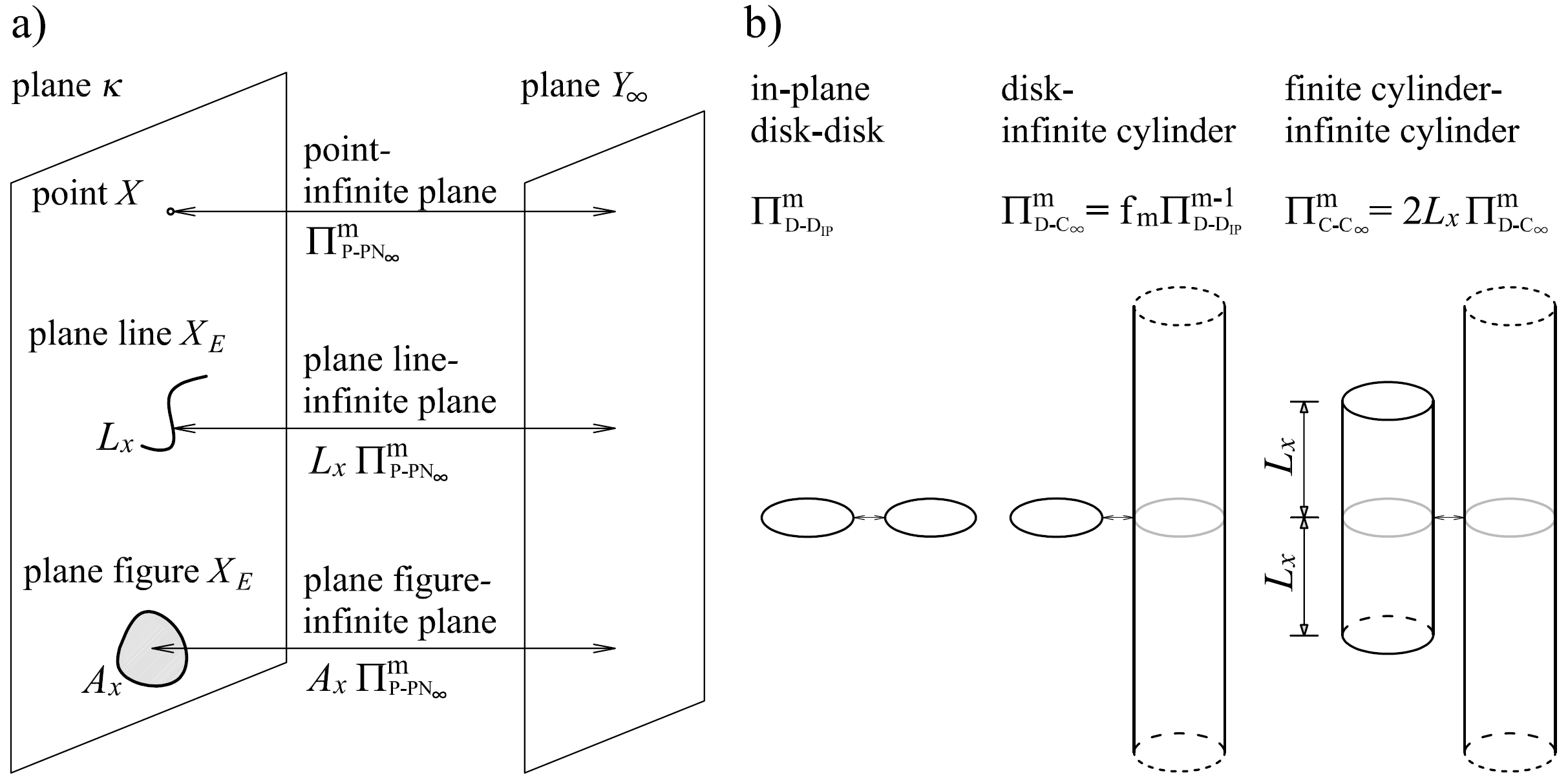}
			\caption{a) Illustration of IUI: The interaction potential between a point $X$ and an infinite plane $Y_\infty$ is $\Pi_{\operatorname{P-PN_{\infty}}}^m$. A plane line and a plane figure are defined in a plane passing through $X$ and parallel to $Y_\infty$. The interaction potential of the plane line or the plane figure with the plane $Y_\infty$ is obtained by multiplying $\Pi_{\operatorname{P-PN_{\infty}}}^m$ with the line's length or the figure's area, respectively. b) Application of RIIPI and IUI: Interaction potential between finite- and infinite-length cylinders in parallel orientation, $\Pi_{\operatorname{C-C_{\infty}}}^m$, equals an interaction potential $\Pi_{\operatorname{D-D_{IP}}}^{m-1}$ between two in-plane disks multiplied by the factor $f_m$ and by the cylinder's length.}
			\label{fig:IUI}
		\end{center}
	\end{figure}

Now we can make one observation regarding the interaction potential between a finite-sized body $X_E$ and the body $Y_{\infty}$ that has one or two infinite dimensions perpendicular to the gap, $\Pi_{X_E-Y_{\infty}}^{m}$. The body $X_E$ is formed such that a geometrical object $X$ (point, line, plane figure, etc.) is extended along the dimensions that are parallel to the infinite dimensions of $Y_{\infty}$. For such bodies, $\Pi_{X_E-Y_{\infty}}^{m}$ represents an interaction potential per unit length or area of object $X_E$. For example, consider an interaction potential between a point $X$ and an infinite plane $Y_\infty$, $\Pi_{\operatorname{P-PN_{\infty}}}^m$, as in Fig.~\ref{fig:IUI}a.
Since the plane $Y_{\infty}$ has two infinite dimensions perpendicular to the gap, let us define a plane $\kappa$ through the point $X$ that is parallel to the plane $Y_{\infty}$. The object $X_E$ is now formed by extending the point $X$ inside the plane $\kappa$, to form a plane line with length $L_x$ or a plane figure with area $A_x$. 
Since the plane $Y_{\infty}$ is infinite, integration of the interaction potential $\Pi_{\operatorname{P-PN_{\infty}}}^m$ over a line or a figure lying in the $\kappa$-plane is trivial and equals the multiplication of $\Pi_{\operatorname{P-PN_{\infty}}}^m$ with $L_x$ or $A_x$, respectively. Therefore, a point-infinite plane interaction can be considered as (i) a plane line-infinite plane interaction per unit length of a line and (ii) a plane figure-infinite plane interaction per unit area of a plane figure. 
Let us designate this observation as \emph{an interaction between a unit and an infinite-dimension body in parallel orientation} (IUI).

As an example, RIIPI and IUI suggest that, to find an interaction of $m^{th}$ order between a finite-length cylinder $X$ and an infinite-length cylinder $Y_{\infty}$ in parallel orientation, one only needs to calculate an interaction potential of $\left(m-1\right)^{th}$ order between two in-plane disks (with appropriate radii), multiply it by the factor $f_m$ and by the length of the cylinder $X$, cf. Fig.~\ref{fig:IUI}b.  We will use these observations to find several general interaction potential laws.

%
%

Next, let us find a point-infinite plate interaction potential by integrating the point-infinite plane law \eqref{eq1rule2} over the $Y$-plate thickness, along $y_1$. The thickness of the $Y$-plate is denoted $2b_y$ and the result is
\begin{equation}
	\begin{aligned}
		\Pi_{\operatorname{P-PT_{\infty 23}}}^{m}&=\int_{x_1+q_2}^{x_1+q_2+2b_y} \Pi_{\operatorname{P-PN_{\infty 23}}}^{m} \dd{y_1} = \frac{2\pi \left[q_2^{3-m} - \left(2b_y+q_2\right)^{3-m}\right]}{\left(m-3\right) \left(m-2\right)}\; \text{for} \; m>3,
	\end{aligned}
	\label{eqP-PT}
\end{equation}
where $q_2$ is the gap between the point and the plate. From this relation, we can derive a point-infinite half-space law, $\operatorname{P-HS_{\infty}}$, by letting $b_y\rightarrow \infty$,
\begin{equation}
	\begin{aligned}
		\Pi_{\operatorname{P-HS_{\infty 23}}}^{m}&=\lim\limits_{b_y \rightarrow\infty}\Pi_{\operatorname{P-PT_{\infty 23}}}^{m} = \frac{2 \pi }{\left(m-3\right) \left(m-2\right) q_2^{m-3}}\; \text{for} \;  m>4.
	\end{aligned}
	\label{eqP-HS}
\end{equation}
The last two expressions will be employed in Subsection \ref{diskplatesub} to derive $\operatorname{D-HS_{\infty}}$ and $\operatorname{D-PT_{\infty}}$ laws. Let us note that the $\operatorname{P-PT_{\infty}}$ law is actually a difference of two $\operatorname{P-HS_{\infty}}$ laws obtained for the two gap values, $q_2$ and $2b_y+q_2$. We will generalize this observation at the end of this section.

Let us return to the P-PN$_{\infty}$ law \eqref{eq1rule2} which is also the unit-area plane figure-infinite plane law. By integrating \eqref{eq1rule2} over the $X$-plate thickness, along $x_1$, a unit-area plate-infinite plane law follows, i.e.
\begin{equation}
	\begin{aligned}
		\Pi_{\operatorname{PT-PN_{\infty 23}}}^{m}&=\int_{-b_x}^{b_x} \Pi_{\operatorname{P-PN_{\infty 23}}}^{m} \dd{x_1} = 2 \pi \frac{ \left(y_1-b_x\right)^{3-m}-\left(y_1+b_x\right)^{3-m}}{\left(m-3\right)\left(m-2\right)}\; \text{for} \; m>3.
	\end{aligned}
	\label{eqPT-PN}
\end{equation}
Here, $y_1$ is the distance between the middle of the plate and the plane, while $2 b_x$ is the thickness of the plate. By integrating the last expression over the $Y$-plate thickness, along $y_1$, we find a plate (thickness $2 b_x$, unit area) - plate (thickness $2 b_y$, infinite area) law, where $q_2$ is the gap between these parallel plates, i.e.
\begin{equation}
	\begin{aligned}
		\Pi_{\operatorname{PT-PT_{\infty 23}}}^{m}&=\int_{q2+b_x}^{q_2+b_x+2b_y} \Pi_{\operatorname{PT-PN_{\infty 23}}}^{m} \dd{y_1} \\
		&= 2 \pi \frac{q_2^{4-m} -  (2 b_x + q_2)^{4-m}-(2 b_y + q_2)^{4-m}+[2(b_x+b_y)+q_2]^{4-m}}{(m-4)(m-3)(m-2)}\; \text{for} \; m>4.
	\end{aligned}
	\label{eqPT-PT}
\end{equation}
For the special case of vdW attraction, this expression reduces to the one given in \cite{2001tadmor}. From this general PT-PT$_{\infty}$ law, we can obtain two special cases: (i) a PT-HS$_{\infty}$ law, by letting $b_y \rightarrow \infty$, i.e.
\begin{equation}
	\begin{aligned}
		\Pi_{\operatorname{PT-HS_{\infty 23}}}^{m}&=\lim_{b_y \rightarrow \infty} \Pi_{\operatorname{PT-PT_{\infty 23}}}^{m} = 2 \pi \frac{q_2^{4-m} -  \left(2 b_x + q_2\right)^{4-m}}{\left(m-4\right)\left(m-3\right)\left(m-2\right)}\; \text{for} \; m>4,
	\end{aligned}
	\label{eqPT-HS}
\end{equation}
and (ii) a HS-HS$_{\infty}$ law, by letting $b_x\rightarrow\infty$ in the last expression, i.e.
\begin{equation}
	\begin{aligned}
		\Pi_{\operatorname{HS-HS_{\infty}}}^{m}&=\lim_{b_x \rightarrow \infty} \Pi_{\operatorname{PT-HS_{\infty}}}^{m} = 2 \pi \frac{q_2^{4-m} }{\left(m-4\right)\left(m-3\right)\left(m-2\right)}\; \text{for} \; m>4,
	\end{aligned}
	\label{eq1rule6}
\end{equation}
which is a well-known expression \cite{2011israelachvili}. 

Let us note that the PT-PT$_{\infty}$ interaction can be obtained from the HS-HS$_{\infty}$ interaction. If we explicitly designate HS-HS$_{\infty}$ law \eqref{eq1rule6} as a function of the gap $a=q_2$, i.e. $ \Pi_{\operatorname{HS-HS_{\infty}}}^{m}(a)$, then the PT-PT$_{\infty}$ interaction is
\begin{equation}
	\begin{aligned}
		\Pi_{\operatorname{PT-PT_{\infty}}}^{m}(q_2,b_x,b_y)&=\Pi_{\operatorname{HS-HS_{\infty}}}^{m}(q_2)+\Pi_{\operatorname{HS-HS_{\infty}}}^{m}(2b_x+2 b_y + q_2) \\
		&- \Pi_{\operatorname{HS-HS_{\infty}}}^{m}(2b_x + q_2) - \Pi_{\operatorname{HS-HS_{\infty}}}^{m}(2 b_y +q_2).
	\end{aligned}
	\label{eq1rule9}
\end{equation}
This observation can be applied in various situations. For example, equations \eqref{eqP-PT} and \eqref{eqP-HS} represent a special case of this rule. Let as designate it as the \emph{rule of infinite half-body interaction} (RIHBI). It is valid only for rectangular geometries in a symmetric configuration, and does not require that either body extends infinitely. Although not explicitly defined, this rule is used, for example, in \cite{1960derocco} to find an in-plane rectangle-rectangle interaction potential.

Finally, from the interactions between an unit-area plate with thickness $b_y$ and either infinite plane, plate or half-space, we can obtain interactions between a rectangular section (with dimensions $2 b_y \times 2 h_x$) with an infinite plane (R-PN$_\infty$), plate (R-PT$_\infty$) or half-space (R-HS$_\infty$) in the orthogonal orientation, by simply multiplying appropriate laws \eqref{eqPT-PN}, \eqref{eqPT-PT}, \eqref{eqPT-HS} with $2 h_x$, respectively. In line with IUI, these R-PN$_\infty$, R-PT$_\infty$ and R-HS$_\infty$ laws are also interaction potentials between the unit-length rectangular prism and an infinite plane, plate or half-space in the parallel orientation. Rectangle-rectangle laws will be considered in Section \ref{rectsection}.

For simplicity, we will remove indices related to direction, plane, and infinity almost everywhere in the following. We will only use them when necessary to avoid ambiguity.

\section{Circular cross sections}
\label{circsec}

First, we consider the interaction between in-plane disks and then introduce an offset between their planes. Both exact and approximate D-D laws are studied. Second, we deal with P-C and D-C interactions, followed with D-HS and D-PT laws.

\subsection{In-plane disk-disk interaction for arbitrary exponent $m$}
\label{inplanedisks}


For the derivation of in-plane disk-disk ($\operatorname{D-D_{IP}}$) laws, we follow two approaches that differ in the utilized relative polar coordinate system \cite{1972langbein}: (i) RPCS1, see Fig.~\ref{fig:intro1}b and (ii) RPCS2, see Fig.~\ref{fig:intro1}c. For both approaches, the distance between point pairs is $r=p$. As an intermediate step to the $\operatorname{D-D_{IP}}$ expressions, we discuss in-plane point-disk (P-D$_{\text{IP}}$) laws.


Using RPCS1, the P-D$_{\text{IP}}$ potential is obtained by integrating the P-P potential over the area of disk $X$, i.e.
\begin{equation}
\label{eq:pdip1}
	\begin{aligned}
		\Pi_{\operatorname{P-D_{IP}}}^m&= 2 \int_{t-R_x}^{t+R_x} \arccos(\frac{p^2+t^2-R_x^2}{2pt})  \frac{1}{p^m} p\dd{p},
	\end{aligned}
\end{equation}
see \cite{1972langbein,2024borkovićd} for details. This integral has analytical solutions for arbitrary integer exponents $m>2$. The expressions for $m=3,4,5,...12$ are given in \hyperref[supsec]{Notebook 2}. To find the $\operatorname{D-D_{IP}}$ laws, we need to integrate P-D$_{\text{IP}}$ over the area of disk $Y$, i.e.
\begin{equation}
	\label{eq:intDD}
	\begin{aligned}
		\Pi_{\operatorname{D-D}}^m&=\int_{A_x} \Pi_{\operatorname{P-D_{IP}}}^m \dd{A_x}= 2 \int _{R_x+q_2}^{R_x+q_2+2 R_y} \Pi_{\operatorname{P-D_{IP}}}^m \arccos(\frac{t^2+d^2-R_y^2}{2td}) t\dd{t}.
	\end{aligned}
\end{equation}
This integration is significantly more involved than Eq.~\eqref{eq:pdip1} and it turns out that exact analytical expressions exist only for even exponents $m>3$, see \hyperref[supsec]{Notebook 2}.

The second approach utilizes RPCS2. With such parameterization, a general expression for the P-D$_{\text{IP}}$ interaction can be found, i.e.
\begin{equation}
	\label{eq:pdip2}
	\begin{aligned}
		\bar{\Pi}_{\operatorname{P-D_{IP}}}^m&= 2 \int_{0}^{R_x} \int_{0}^{\pi} \frac{1}{p^m} r_x \dd{\varphi_x} \dd{r_x} =\pi R_x^2 \, t^{-m}\, _2F_1 \left(\frac{m}{2},\frac{m}{2};2;\frac{R_x^2}{t^2}\right) \; \text{for} \; m>2,
	\end{aligned}
\end{equation}
see Appendix \ref{appendix} for a detailed derivation. Here, $_2F_1 \left(a,b;c;z\right)$ is the Gaussian hypergeometric function
\begin{equation}
	\label{eq: ip13hyp}
	\begin{aligned}
		_2F_1 \left(a,b;c;z\right)= \sum_{k=0}^{\infty} \frac{\left(a\right)_k \left(b\right)_k}{\left(c\right)_k} \frac{z^k}{k!}
	\end{aligned}
\end{equation}
and $\left(a\right)_k$ is the Pochhammer symbol
\begin{equation}
	\label{eq: ip13poch}
	\begin{aligned}
		\left(a\right)_k=\frac{\Gamma \left(a+k\right)}{\Gamma \left(k\right)}.
	\end{aligned}
\end{equation}
By replacing the exponent $m$ with integer values, expression \eqref{eq:pdip2} evaluates to rational functions for even $m$, while the expressions are much more complicated for odd $m$ and consist of either elliptic integrals or hypergeometric functions, depending on the representation. Nevertheless, it is interesting that WM14 does not recognize that the expressions obtained with \eqref{eq:pdip1} and \ref{eq:pdip2} are the same for odd $m$, while it does for even $m$, cf. \hyperref[supsec]{Notebook 2}. When considering the numerical evaluation, these expressions are in full agreement for an arbitrary integer $m>2$. Finally, we need to integrate Eq.~\eqref{eq:pdip2} over the area of disk $Y$ to find the $\operatorname{D-D_{IP}}$ law, and the result is the infinite series
\begin{equation}
	\label{eqsingleser}
	\begin{aligned}
		\bar{\Pi}_{\operatorname{D-D_{IP}}}^{m} = \frac{\pi^2 R_x^2 R_y^2}{d^{m}\Gamma^2(\frac{m}{2})}    \sum_{n=0}^{\infty} \frac{\Gamma^2 (n+\frac{m}{2})}{\Gamma (n+1) \Gamma (n+2)} \left(\frac{R_y}{d}\right)^{2n} \, _2F_1 \left(n+\frac{m}{2},n+\frac{m}{2};2;\frac{R_x^2}{d^2}\right)\; \text{for} \; m>\frac{7}{2},
	\end{aligned}
\end{equation}
see Appendix \ref{appendix}. Since WM14 calculates the $_2F_1$ function efficiently, evaluation of expression \eqref{eqsingleser} is more efficient that the double series solution in \cite{1972langbein}.

For the special case of equal radii $R_x=R_y=R$, the series \eqref{eqsingleser} can be represented as
\begin{equation}
	\begin{aligned}
		\tilde{\Pi}_{\operatorname{D-D_{IP}}}^{m} &= \frac{\pi^2 R^4}{d^{m}} \, _pF_q \left(\frac{3}{2},\frac{m}{2},\frac{m}{2};2,3;\frac{4 R^2}{d^2} \right)\; \text{for} \; m>\frac{7}{2},
	\end{aligned}
	\label{eq:DDIPspec}
\end{equation}
where $_pF_q$ is the generalized hypergeometric function, i.e.
\begin{equation}
	\begin{aligned}
		 _pF_q (a_1, \dotsc,a_p;b_1, \dotsc, b_q;z) = \sum_{k=0}^{\infty} \frac{(a_1)_k (a_2)_k \cdots (a_p)_k}{(b_1)_k (b_2)_k \cdots (b_q)_k} \frac{z^k}{k!}.
	\end{aligned}
\end{equation}
To the best of our knowledge, this is the first representation of the $\operatorname{D-D_{IP}}$ law for equal radii and general $m$ in the form of a generalized hypergeometric function. In an elegant manner, function $\tilde{\Pi}_{\operatorname{D-D_{IP}}}^{m}$ simplifies to a rational function for even $m$, cf. \hyperref[supsec]{Notebook 2}. For example, by taking $m=6$, a correct expression is obtained \cite{1939dube}. For odd $m$, the result is in the form of a generalized hypergeometric function or elliptic integrals.

\subsection{Exact disk-disk law for vdW attraction}
\label{secexactDD}

When there is an offset $q_1$ between the disks, the distance between point pairs is $r=\left(p^2+q_1^2\right)^{1/2}$. Here, we now consider only vdW attraction since analytical solutions for an arbitrary $m$ are quite difficult or even impossible to obtain. So, for $m=6$, the interaction potential between two disks, using RPCS1, is
\begin{equation}
	\label{eq:ddset}
	\begin{aligned}
		\Pi_{\operatorname{D-D}}^m&=\int_{A_x} \int_{A_y} \frac{1}{r^6}\dd{A_y}\dd{A_x} =\int_{A_x} \int_{A_y} \left(p^2+q_1^2\right)^{-3} \dd{A_y}\dd{A_x}\\
		&= 2 \int _{R_x+q_2}^{R_x+q_2+2 R_y}  \Pi_{\operatorname{P-D}}^6 \arccos(\frac{t^2+d^2-R_y^2}{2td}) t \dd{t},
	\end{aligned}
\end{equation}
where the potential between a point and a disk is
\begin{equation}
	\label{eq:pd}
			\begin{aligned}
					\Pi_{\operatorname{P-D}}^6&=2  \int_{t-R_x}^{t+R_x} \arccos(\frac{p^2+t^2-R_x^2}{2pt})  \frac{1}{\left(p^2+q_1^2\right)^3} p\dd{p}.
			\end{aligned}
\end{equation}
As in Subsection \ref{inplanedisks}, the first step is to find the P-D law by solving \eqref{eq:pd}. The resulting primitive function has a discontinuity at $p=\frac{t^2-R_x^2}{\sqrt{t^2+R_x^2}}$ and we must integrate it carefully. The resulting expression has a singularity for $q_1=0$, but we can fix it by multiplying the numerator and denominator with appropriate factors. The final result is the vdW P-D law (with an offset)
\begin{equation}
	\begin{aligned}
	\Pi_{\operatorname{P-D}}^6&=\frac{N_6}{\rho ^{3} D_6},
	\end{aligned}
\end{equation}
where
\begin{equation}
	\begin{aligned}
		N_6&\coloneqq \pi  R_x^2 \left(2
		q_1^6+q_1^4 \left(5
		R_x^2+6 t^2\right)+q_1^2 \left(6
		R_x^2 t^2+4 R_x^4+6 t^4\right)+2
		t^6-3 R_x^2 t^4+R_x^6\right)\\
		D_6& \coloneqq q_1^4
		\left(\rho-R_x^2+3
		t^2\right)+q_1^2
		\left(R_x^2+t^2\right) \left(2
		\rho-3 R_x^2+3
		t^2\right)+\left(R_x^2-t^2\right)^2
		\left(\rho-R_x^2+t^2\right)+q_1^6 \\
		\rho& \coloneqq \sqrt{q_1^4+ 2 q_1^2 \left(R_x^2+t^2\right)+\left(R_x^2-t^2\right)^2}.
	\end{aligned}
\end{equation}

The second step, solving integral \eqref{eq:ddset}, is more involving and the primitive function consists of elliptic integrals. The lower limit is pleasantly simple, while the upper one is significantly more complicated, see \hyperref[supsec]{Notebook 3}. The resulting vdW D-D law becomes
\begin{equation}
	\label{DDlaw}
	\begin{aligned}
		\Pi_{\operatorname{D-D}}^6&= \frac{2 \pi^2 R_y^2}{8 q_1^4} +\frac{\pi}{8 q_1^4\sqrt{a_1}} \left\{ (a_1+a_7) K (\tfrac{a_5}{a_1}) -a_1 E (\tfrac{a_5}{a_1}) +2(q_1-i q_2) (d_4 - i q_1) \right.  \\
		&	\left.  \left[i\left(R_x^2-2R_y^2\right)\mathit{\Pi} (\tfrac{4 a_4}{a_3},\tfrac{a_5}{a_1})+\frac{R_x^2\left(i q_1+R_x\right)d_1 \mathit{\Pi}(\tfrac{ a_6}{a_2},\tfrac{a_5}{a_1})}{\left(q_1+i R_x\right)d_3} \right] \right\},
	\end{aligned}
\end{equation}
where
\begin{equation}
	\label{eq:DDlawDesig1}
	\begin{aligned}
		a_1 &\coloneqq \left(q_1+iq_2\right) \left(q_1-i d_5\right) \left(q_1^2+2 i q_1 R_y +d_2 d_4\right) \\
		a_2 &\coloneqq d_3^2 \left(iq_1+d_5\right) \left(d_2-i q_1\right) \\
		a_3 &\coloneqq q_1^2 + 2 i q_1 R_x + d_2 d_5 \\
		a_4 &\coloneqq R_y d \\
		a_5 &\coloneqq -16 i a_4 q_1 R_x \\
		a_6 &\coloneqq -4 a_4 \left(q_1 + i R_x\right)^2 \\
		a_7 &\coloneqq - \frac{4 q_1 R_x \left[q_1^2 R_x + R_x d_2d_5 + 2iq_1 \left(\left(R_x-2R_y\right) \left(R_x+R_y\right) -q_2R_y\right)  \right]}{q_1+i R_x} 
	\end{aligned}
\end{equation}
and
\begin{equation}
	\label{eq:DDlawDesig2}
	\begin{aligned}
		d&\coloneqq q_2+R_x+R_y,\quad d_1 \coloneqq q_2 + R_x + 2  R_y,  \quad d_2 \coloneqq q_2 + 2  (R_x + R_y) \\
		d_3 & \coloneqq q_2 + R_x, \quad d_4 \coloneqq q_2 + 2 R_x, \quad d_5 \coloneqq q_2 + 2  R_y.
	\end{aligned}
\end{equation}
Here $K$, $E$, and $\mathit{\Pi}$ are the complete elliptic integrals of the first, second, and third kind. The obtained expression is correct but has complex-valued arguments that do not allow taking the real part of the expression analytically. Additionally, the expression cannot be integrated w.r.t.~$q_1$, the limit $q_1 \rightarrow 0$ does not exist, and the evaluation for $q_1 \approx 0$ requires an arbitrary precision arithmetic. 

On the other hand, the component of the interaction force along the gap, $\partial \Pi_{\operatorname{D-D}}^6 / \partial q_2$, is much simpler than the potential, since the complete elliptic integrals of the third kind cancel out. This force component has a proper limit for $q_1 \rightarrow 0$ that is perfectly aligned with the results obtained in Subsection \ref{inplanedisks} for $\operatorname{D-D_{IP}}$, cf. \hyperref[supsec]{Notebook 3}. This agreement additionally confirms that the derived D-D law in  \eqref{DDlaw} is accurate.


\subsection{Approximate disk-disk law for even exponent $m$}
\label{secappDD}

An approximate disk-disk law named \emph{improved section-section interaction potential} (ISSIP), which provides good accuracy for small separations, is derived in \cite{2024borkovićd}. It is represented as a product of an approximate D-D\textsubscript{IP} law and a hypergeometric function, i.e.
\begin{equation}
	\label{appDD1}
	\begin{aligned}
		\Pi_{\mathrm{ISSIP}} &= \hat{\Pi}_{\operatorname{D-D_{IP}}}^m \, _2F_1 \left(\frac{2m-7}{4},\frac{2m-5}{4};\frac{m}{2};-\frac{q_1^2}{q_2^2}\right), \\
		 \hat{\Pi}_{\operatorname{D-D_{IP}}}^m &= 2^{\frac{5}{2}-m} \pi^\frac{3}{2} \sqrt{\frac{R_x R_y}{R_x + R_y}} \frac{\Gamma (m-\frac{7}{2})}{\Gamma (\frac{m}{2})^2} q_2^{\frac{7}{2}-m}.
	\end{aligned}
\end{equation}
For even $m$, this law evaluates to elliptic integrals of the first and second kind, while for odd $m$ we obtain rational functions. 

\begin{figure}[h!]
	\centering
	\begin{tikzpicture}
		\begin{axis}[
			xlabel = {Log $q_2$},
			ylabel = {Scaling factor},
			ylabel near ticks,
			legend pos=south east,
			legend cell align=left,
			legend style={font=\scriptsize},
			width=0.5\textwidth,
			minor y tick num = 1,
			minor x tick num = 2,
		    ytick distance=1,
			xtick distance=4,
			clip=false,grid=both];
			\node [text width=1em,anchor=north west] at (rel axis cs: -0.195,1.1){\subcaption{\label{fig:a}}};
			\addplot[green,thick] table [col sep=comma] {data/dataDDscalingEx1.csv};
			\addplot[black,thick,dashed] table [col sep=comma] {data/dataDDscalingEx2.csv};
			\addplot[red,dashed,thick] table [col sep=comma] {data/dataDDscalingEx3.csv};
			\addplot[purple,thick] table [col sep=comma] {data/dataDDscalingEx4.csv};
			\addplot[blue,dotted,thick] table [col sep=comma] {data/dataDDscalingEx5.csv};
			\legend{c=1/100,c=1/10,c=1,c=10,c=100} 
		\end{axis}
	\end{tikzpicture}
	\begin{tikzpicture}
		\begin{axis}[
			xlabel = {Log $q_2$},
			ylabel = {Scaling factor},
			ylabel near ticks,
			legend pos=south east,
			legend style={font=\scriptsize},
			legend cell align=left,
			width=0.5\textwidth,
		minor y tick num = 1,
		minor x tick num = 2,
		ytick distance=1,
		xtick distance=4,
			clip=false,grid=both]
			\node [text width=1em,anchor=north west] at (rel axis cs: -0.195,1.1){\subcaption{\label{fig:b}}};
			\addplot[green,thick] table [col sep=comma] {data/dataDDscalingApp1.csv};
			\addplot[red,dashed,thick] table [col sep=comma] {data/dataDDscalingApp2.csv};
			\addplot[blue,dotted,thick] table [col sep=comma] {data/dataDDscalingApp3.csv};
			\legend{c=1/100,c=1,c=100} 
		\end{axis}
	\end{tikzpicture}
	\caption{Scaling of the vdW disk-disk laws for different ratios $c=q_1/q_2$ ($R_x=R_y=1$). a) Exact D-D law from Eq.~\eqref{DDlaw}. b) Approximate $\operatorname{D-D_{app}}$ law from Eq.~\eqref{appDD2}. }
	\label{fig:LJpoten3}
\end{figure}
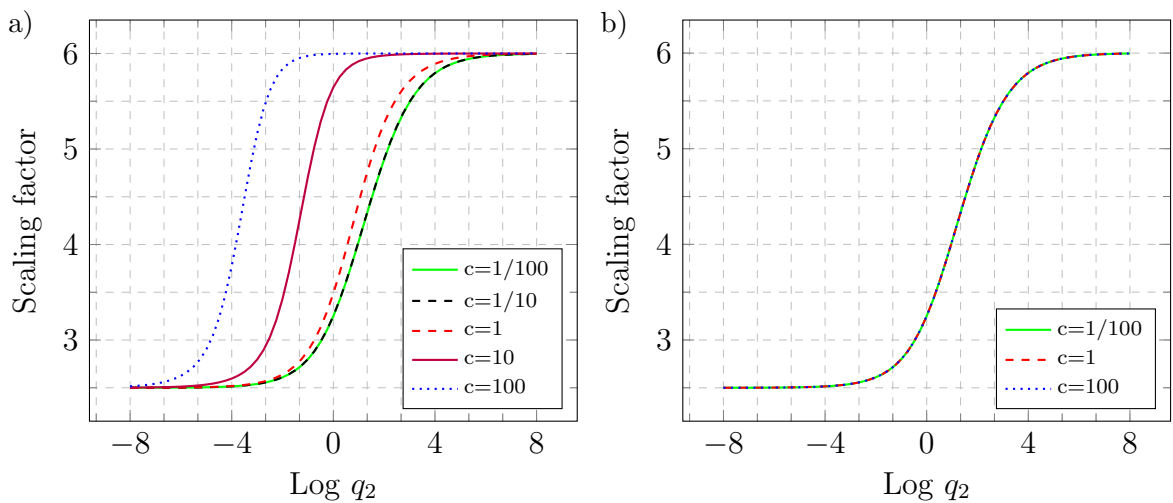

Here, we propose to further improve the ISSIP law by replacing the approximate potential $\hat{\Pi}_{\operatorname{D-D_{IP}}}^m$ in \eqref{appDD1} with the exact one for even $m$, see Subsection \ref{inplanedisks}. Therefore, the new approximate disk-disk law, $\operatorname{D-D_{app}}$, is defined as
\begin{equation}
	\label{appDD2}
	\begin{aligned}
		\Pi_{\operatorname{D-D_{app}}}^m = \Pi_{\operatorname{D-D_{IP}}}^m \, _2F_1 \left(\frac{2m-7}{4},\frac{2m-5}{4};\frac{m}{2};-\frac{q_1^2}{q_2^2}\right)\; \text{for} \; m=4,6,8,...
	\end{aligned}
\end{equation}
Let us compare the exact and the approximate D-D laws for $m=6$. The scaling factor function of the potential vs.~the gap is shown in Fig.~\ref{fig:LJpoten3} for different values of ratio $c=q_1/q_2$.
For the D-D law and $c>1/10$, the transition region between small and large separations strongly depends on ratio $c$, see Fig.~\ref{fig:LJpoten3}a. The rate of change between small and large separation limits increases with $c$. Regarding the $\operatorname{D-D_{app}}$ law, it returns proper asymptotic scaling, 5/2 for small and 6 for large separations. However, due to approximations in $\operatorname{D-D_{app}}$, the scaling factor is practically insensitive to $c=q_1/q_2$, cf. Fig.~\ref{fig:LJpoten3}b. The scaling of the $\operatorname{D-D_{app}}$ law mainly corresponds to the scaling of the $\operatorname{D-D_{IP}}$ law, which can be concluded from Fig.\ref{fig:LJpoten3}a where the results are the same for $c \le 1/10$. It should be emphasized that the previously used ISSIP law \cite{2024borkovićd} is not capable of modeling correct scaling at moderate and large separations.

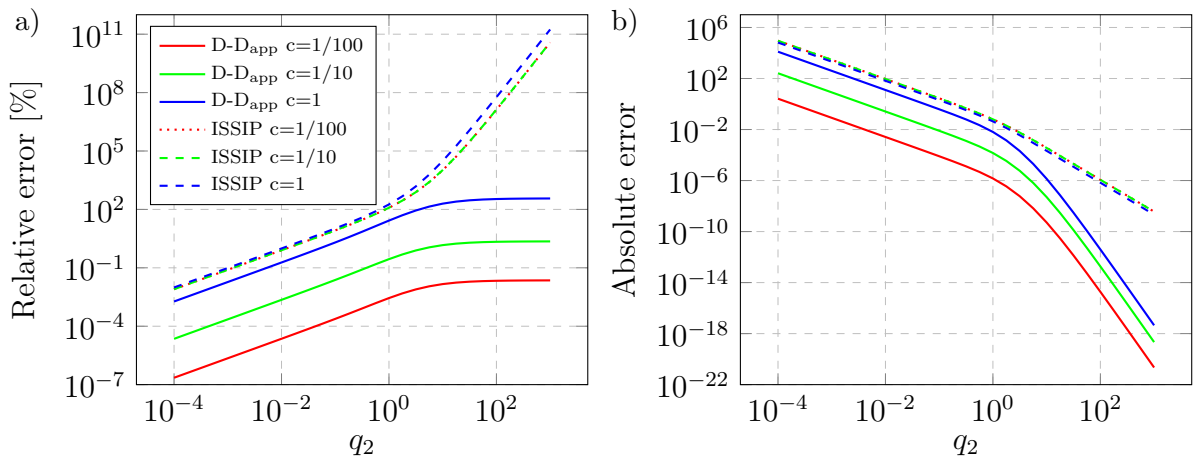
\begin{figure}[h!]
	\centering
	\begin{tikzpicture}
		\begin{loglogaxis}[
			xlabel = {$q_2$},
			ylabel = {Relative error $[\%]$},
			ylabel near ticks,
			legend pos=north west,
			legend cell align=left,
			legend style={font=\scriptsize,nodes={scale=0.91, transform shape}},
			width=0.47\textwidth,
			yminorticks=true,
			grid=major,
			ymin = 10^(-7), ymax = 10^(12),
			ytick distance=10^3,
			xtick distance=10^2,
			clip=false
			];
			\node [text width=1em,anchor=north west] at (rel axis cs: -0.3,1.1){\subcaption{\label{fig:a}}};
			\addplot[red,thick] table [col sep=comma] {data/dataDDAppvsISSIP1.csv};
			\addplot[green,thick] table [col sep=comma] {data/dataDDAppvsISSIP2.csv};
			\addplot[blue,thick] table [col sep=comma] {data/dataDDAppvsISSIP3.csv};
			\addplot[red,dotted,thick] table [col sep=comma] {data/dataDDAppvsISSIP4.csv};
			\addplot[green,dashed,thick] table [col sep=comma] {data/dataDDAppvsISSIP5.csv};
			\addplot[blue,dashed,thick] table [col sep=comma] {data/dataDDAppvsISSIP6.csv};
			\legend{$\operatorname{D-D_{app}}$ c=1/100,$\operatorname{D-D_{app}}$ c=1/10, $\operatorname{D-D_{app}}$ c=1,ISSIP c=1/100,ISSIP c=1/10, ISSIP c=1} 
		\end{loglogaxis}
	\end{tikzpicture}
	\begin{tikzpicture}
		\begin{loglogaxis}[
			xlabel = {$q_2$},
			ylabel = {Absolute error},
			ylabel near ticks,
			legend pos=south west,
			legend cell align=left,
			legend style={font=\scriptsize,nodes={scale=0.95, transform shape}},
			width=0.47\textwidth,
			yminorticks=true,
			grid=major,
			ymin = 10^(-22), ymax = 10^(7),
			clip=false,
			ytick distance=10^4
			];
			\node [text width=1em,anchor=north west] at (rel axis cs: -0.315,1.1){\subcaption{\label{fig:b}}};
			\addplot[red,thick] table [col sep=comma] {data/dataDDAppvsISSIPAbsError1.csv};
			\addplot[green,thick] table [col sep=comma] {data/dataDDAppvsISSIPAbsError2.csv};
			\addplot[blue,thick] table [col sep=comma] {data/dataDDAppvsISSIPAbsError3.csv};
			\addplot[red,dotted,thick] table [col sep=comma] {data/dataDDAppvsISSIPAbsError4.csv};
			\addplot[green,dashed,thick] table [col sep=comma] {data/dataDDAppvsISSIPAbsError5.csv};
			\addplot[blue,dashed,thick] table [col sep=comma] {data/dataDDAppvsISSIPAbsError6.csv};
		\end{loglogaxis}
	\end{tikzpicture}
	\caption{Error of the ISSIP and $\operatorname{D-D_{app}}$ laws for $m=6$ and different ratios $c=q_1/q_2$ ($R_x=R_y=1$): a) relative error, b) absolute error. $\operatorname{D-D_{app}}$ is much more accurate than ISSIP, especially for large separations, but it still shows errors at large separations compared to the exact D-D law.}
	\label{fig:DDAppvsISSIPcomp}
\end{figure}

Now, let us scrutinize the accuracy of these two approximations. Both relative and absolute errors of the ISSIP and $\operatorname{D-D_{app}}$ laws w.r.t.~the exact D-D law are plotted in Fig.~\ref{fig:DDAppvsISSIPcomp} for different ratios $c=q_1/q_2$. 
For small separations, both laws provide reasonable accuracy, but the $\operatorname{D-D_{app}}$ law is more accurate. The ratio $c=q_1/q_2$ strongly affects the accuracy of the $\operatorname{D-D_{app}}$ law, and only slightly affects the accuracy of the ISSIP law. For large separations, the relative error of the ISSIP law blows up, while the relative error of the $\operatorname{D-D_{app}}$ law is bounded. Regarding the absolute error, it decreases monotonically for the ISSIP law, which corresponds to the blow-up of the relative error for large separations. For the $\operatorname{D-D_{app}}$ law, the absolute error reduces with an increased rate for $q_2>1$, which corresponds to the bounded values of relative errors for large separations.

We will employ the new approximate $\operatorname{D-D_{app}}$ law \eqref{appDD2} in Section \ref{secapp} to numerically model the Lennard-Jones interaction between two deformable fibers, and compare the results with the ISSIP law.

\begin{remark}
	\normalfont
	The scaling factor represents the rate of change of the potential $\Pi (a)$ w.r.t.~the gap $a$ in the $\log-\log$ space. Therefore, the scaling factor function $S(a)$ is
	\begin{equation}
		\label{eqscaling}
		\begin{aligned}
			S(a) := - \frac{d }{d \bar{a}}\log \Pi(\bar{a}) \bigg\rvert_{ \bar a = \log a} \quad \text{with} \quad a=e^{\bar a}.
		\end{aligned}
	\end{equation}
	Since the interaction potentials are here defined as inverse-power laws w.r.t.~the distance, the rate of change of an interaction potential w.r.t.~the gap is negative. To be consistent with the definition of the exponent $m$ in Eq.~\eqref{eq0}, we use the minus sign in front of the scaling function $S(a)$ in Eq.~\eqref{eqscaling}.
\end{remark}

\begin{remark}
	\normalfont
	For the limit of large separations, the interaction between arbitrary finite-sized bodies transforms into the point-point interaction. Therefore, the asymptotic scaling value between finite-sized bodies for large separations is always $m$. For each integration over interacting bodies from $-\infty$ to $\infty$, the asymptotic scaling factor is reduced by a value of $m_r=1$. Regarding the scaling at the limit of small separations, each interacting body can be considered infinite as the gap between them closes. For each integration, the asymptotic scaling factor reduces by a value $m_r$ that depends on the orientation and the shape of interacting bodies. Finding a general rule for the value of $m_r$ in the limit of small separations is not trivial.
\end{remark}

	\begin{remark}
		\normalfont
		If the offset $q_1$ between disks is fixed, then the interaction potential converges to a finite value for $q_2 \rightarrow 0$, and the scaling factor for the limit of small separations is zero. Because of this, we consider the scaling behavior of section-section laws for fixed ratios $c=q_1/q_2$, cf. Fig.~\ref{fig:LJpoten3}.
\end{remark}

\subsection{Point-cylinder interaction}
\label{secPC}

A potential between a point and an infinite cylinder is often derived as an intermediate step for deriving interactions between various bodies and a cylinder, \cite{2000montgomerya,2003kirsch}. Let us consider an $m^{th}$ order interaction between a point $Y$ and an infinite cylinder $X$, i.e.
\begin{equation}
	\begin{aligned}
		\Pi_{\operatorname{P-C}}^m&=\int_{A_x} \int_{-\infty}^\infty \frac{1}{r^m}  \dd{q_1}\dd{A_x} =\int_{A_x} \int_{-\infty}^\infty \left(p^2+q_1^2\right)^{-m/2}  \dd{q_1}\dd{A_x} \\
		&=\frac{\sqrt{\pi } \Gamma \left(\frac{m-1}{2}\right)}{\Gamma \left(\frac{m}{2}\right)} \int_{A_x}  \frac{1}{p^{m-1}} \dd{A_x} =	f_m \Pi_{\operatorname{P-D_{IP}}}^{m-1}. 
	\end{aligned}
	\label{eq:PC}
\end{equation}
In line with RIIPI, a P-C potential of $m^{th}$ order, $\Pi_{\operatorname{P-C}}^m$, can be represented as a product of a factor $f_m$ and a P-D$_{\text{IP}}$ potential of $(m-1)^{th}$ order, $\Pi_{\operatorname{P-D_{IP}}}^{m-1}$. Therefore, for a P-C interaction with even exponent $m$, we require a P-D$_{\text{IP}}$ with odd exponent $m-1$, and vice versa. The P-D$_{\text{IP}}$ expressions are already discussed and derived in Subsection \ref{inplanedisks}. The main issue here is that the P-D$_{\text{IP}}$ laws with odd exponents consist of elliptic integrals, see Subsection \ref{inplanedisks}. Such a form of the P-D$_{\text{IP}}$ laws prevents us from finding the exact D-C laws for even $m$ and the exact D-D$_{\text{IP}}$ laws for odd $m$, since these elliptic integrals cannot be integrated further analytically. This issue motivates us to find approximate P-C laws for even exponents $m>3$. 

\begin{remark}
	\normalfont
	 For the cases of even exponents $m$, the expressions for $\Pi_{\operatorname{P-D_{IP}}}^{m}$ and $\Pi_{\operatorname{D-D_{IP}}}^{m}$ are rational functions and the P-C and D-C laws with odd exponents are pleasantly simple. 
\end{remark}

\begin{remark}
	\normalfont
	We can find different forms of the exact P-C laws, depending on the order of integration and the modeling of a cylinder geometry. For example, we have successfully modeled an infinite cylinder as an infinite set of hemispheres. Although different in form, all these expressions return the same numerical values. 
\end{remark}



\begin{figure}[h!]
	\begin{center}
		\includegraphics[width=0.8\linewidth]{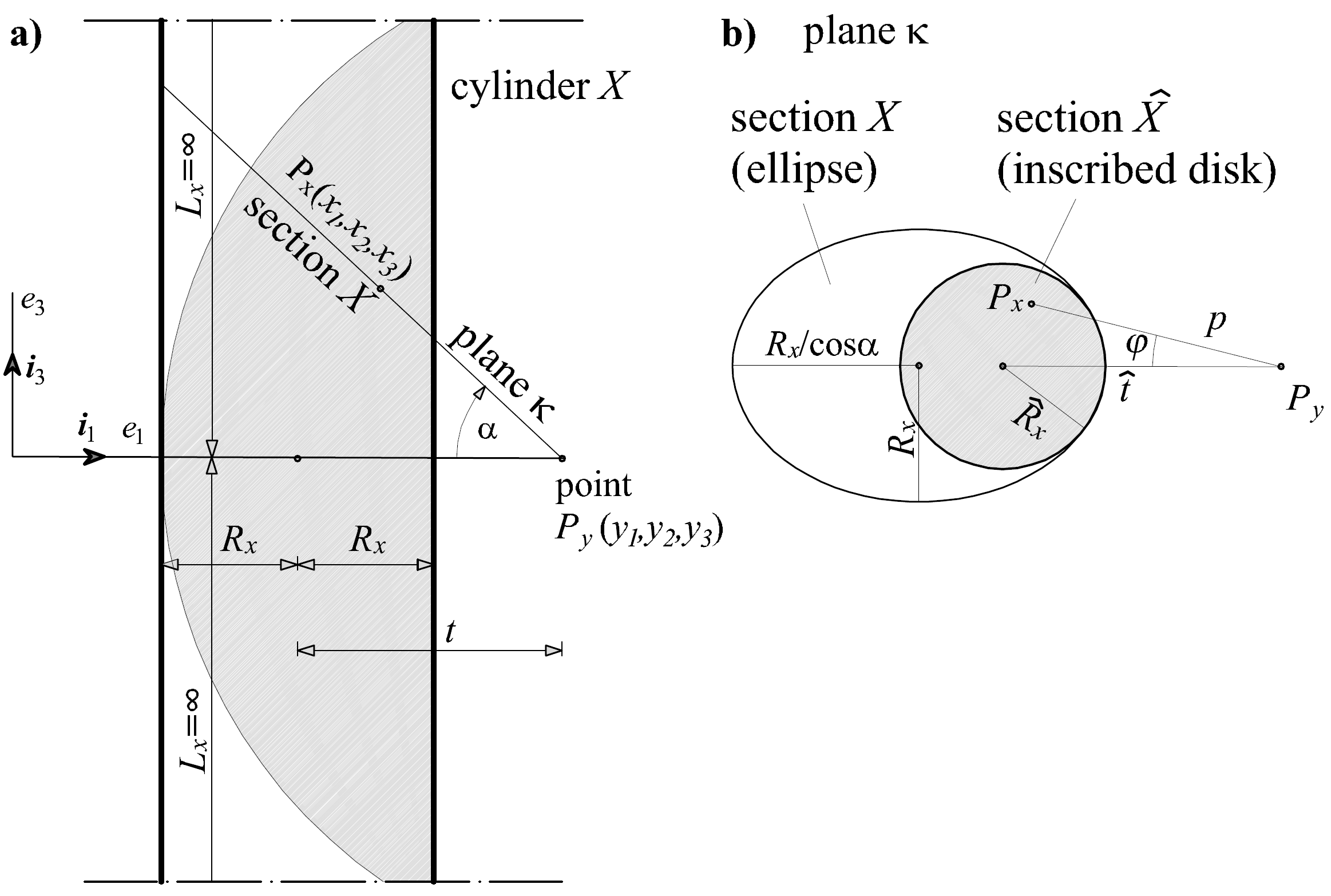}
		\caption{Approximate interaction of a point $P_y$ and an infinite cylinder $x$. a) Front view. b) $\kappa$ plane}
		\label{fig:pc}
	\end{center}
\end{figure}

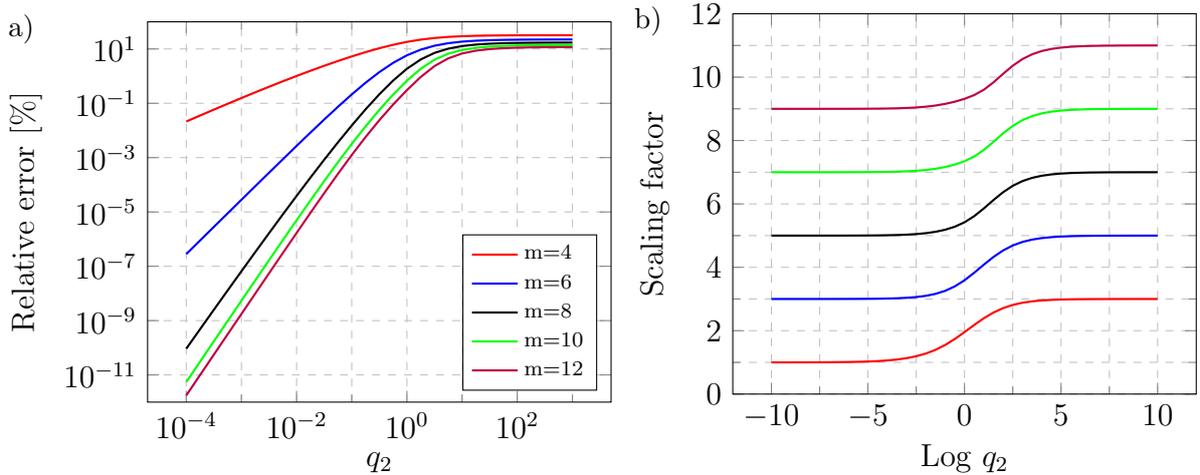
\begin{figure}[h]
	\centering
	\begin{tikzpicture}
		\begin{loglogaxis}[
			xlabel = {$q_2$},
			ylabel = {Relative error $\left[\%\right]$},
			ylabel near ticks,
			legend pos=south east,
			legend cell align=left,
			legend style={font=\scriptsize},
			width=0.48\textwidth,
			ymin = 10^(-12), ymax = 10^2,
			ytick distance=10^2,
			minor xtick={10^(-3),10^(-1),10^1,10^3},
			grid=both,
			clip=false];
			\node [text width=1em,anchor=north west] at (rel axis cs: -0.33,1.1){\subcaption{\label{fig:a}}};
			\addplot[red,thick] table [col sep=comma] {data/dataPCError1.csv};
			\addplot[blue,thick] table [col sep=comma] {data/dataPCError2.csv};
			\addplot[black,thick] table [col sep=comma] {data/dataPCError3.csv};
			\addplot[green,thick] table [col sep=comma] {data/dataPCError4.csv};
			\addplot[purple,thick] table [col sep=comma] {data/dataPCError5.csv};
			\legend{m=4,m=6,m=8,m=10,m=12} 
		\end{loglogaxis}
	\end{tikzpicture}
	\begin{tikzpicture}
		\begin{axis}[
			xlabel = {Log $q_2$},
			ylabel = {Scaling factor},
			ylabel near ticks,
			legend pos=south east,
			width=0.48\textwidth,
			ymin = -0, ymax = 12,
			minor y tick num = 1,
			minor x tick num = 1,
			grid=both,
			clip=false]
			\node [text width=1em,anchor=north west] at (rel axis cs: -0.24,1.1){\subcaption{\label{fig:b}}};
			\addplot[red,thick] table [col sep=comma] {data/dataPCScaling1.csv};
			\addplot[blue,thick] table [col sep=comma] {data/dataPCScaling2.csv};
			\addplot[black,thick] table [col sep=comma] {data/dataPCScaling3.csv};
			\addplot[green,thick] table [col sep=comma] {data/dataPCScaling4.csv};
			\addplot[purple,thick] table [col sep=comma] {data/dataPCScaling5.csv};
		\end{axis}
	\end{tikzpicture}
	\caption{Approximate P-C$_{\text{app}}$ laws for even $m$ and $R_x=1$. a) Relative error. b) Scaling factor. }
	\label{fig:PCerror}
\end{figure}

Let us derive an approximate P-C law. A point $P_y$ and an infinite cylinder $X$ are sketched in Fig.~\ref{fig:pc}a. 
The distance between the point and the cylinder axis is $t$. Let us define a plane $\kappa$ that goes through the point $P_y$ and is perpendicular to the plane $e_1 e_3$. In general, the section between the plane $\kappa$ and the cylinder $X$ has an elliptical shape, see Fig.~\ref{fig:pc}b, and the P-C interaction can be obtained by integrating the point-ellipse interaction over $\alpha \in \left[-\pi/2,\pi/2\right]$. However, this integration results in elliptic integrals and so we pursue the following approximation: Instead of considering an interaction between a point and an elliptical section, let us approximate the elliptical section with an inscribed disk at the closest point, cf. Fig.~\ref{fig:pc}b. The radius of the inscribed disk is $\hat{R_x}$ and the distance between the center of the inscribed disk and $P_y$ is $\hat{t}$. The relation between these quantities and the radius $R_x$ and distance $t$ are
\begin{equation}
	\begin{aligned}
		\hat{R_x} = R_x\cos{\alpha}, \quad \hat{t} = \frac{t-R_x\left(\cos^2{\alpha} -1\right)}{\cos{\alpha}}.
	\end{aligned}
\end{equation}
Therefore, we are performing a volume integration w.r.t.~the coordinates $\left(p, \varphi, \alpha \right)$, see \hyperref[supsec]{Notebook 4}. The Jacobian of the coordinate transformation is $p^2 \cos{\varphi}$ so that the integral to calculate becomes  
\begin{equation}
	\begin{aligned}
		\tilde{\Pi}_{\operatorname{P-C_{app}}}^m &= 2 \int_{0}^{\pi/2} \int_{\hat{t}-\hat{R}_x}^{\hat{t}+\hat{R}_x} \int_{-\hat{\varphi}}^{\hat{\varphi}} \frac{1}{p^m} p^2 \cos{\varphi} \dd{\varphi}\dd{p} \dd{\alpha}, \quad \hat{\varphi} = \arccos(\frac{p^2+\hat{t}^2-\hat{R}_x^2}{2p\hat{t}}).
	\end{aligned}
\end{equation}
WM14 can solve this integral for specific values of integer $m$ and the results are approximate P-C$_{\text{app}}$ laws for $m=4,6,...$, cf. \hyperref[supsec]{Notebook 4}. These expressions are rational functions of $arctanh(t)$ which makes them convenient to evaluate. The error w.r.t.~the numerical integration and the scaling factor of these laws are displayed in Fig.~\ref{fig:PCerror}.
For small separations, the error is very small and decreases with an increase in $m$. The error increases with the gap $q_2$ but has a horizontal asymptote at approximately 30 \%. The asymptotic scaling is correct, $m-3$ for small and $m-1$ for large separations. These results suggest that our P-C$_{\text{app}}$ laws have all required properties, see Section \ref{secintro}. For example, by setting $m=6$, the approximate vdW P-C$_{\text{app}}$ law is
\begin{equation}
	\begin{aligned}
		\tilde{\Pi}_{\operatorname{P-C_{app}}}^6 &= \frac{\pi}{64 (R_x+t)^2} \left[\frac{  (11 R_x +5t) \text{arctanh}\sqrt{\frac{2 R_x}{R_x+t}}}{ \sqrt{2\left(R_x+t\right)}  R_x^{3/2}} - \frac{ 83 R_x^3 + 43 R_x^2 t + 17 R_x t^2 -15 t^3 }{3 R_x \left(R_x-t\right)^3 }\right].
	\end{aligned}
\end{equation}

\subsection{Disk-cylinder interaction}

When considering an interaction between a disk and an infinite cylinder in parallel orientation, the situation is similar to the P-C case, since the initial integration can be done w.r.t.~offset $q_1$ and we can apply RIIPI to the $\operatorname{D-D_{IP}}$ law. Also, we can start from the P-C laws and integrate over the cross section of a disk. This gives a potential between a disk (or a unit-length cylinder) and an infinite cylinder,
\begin{equation}
	\begin{aligned}
		\Pi_{\operatorname{D-C}}^m&=\int_{A_y} \Pi_{\operatorname{P-C}}^m \dd{A_y} = f_m \int_{A_y} \Pi_{\operatorname{P-D_{IP}}}^{m-1} \dd{A_y} =	f_m \Pi_{\operatorname{D-D_{IP}}}^{m-1}.
	\end{aligned}
\end{equation}
According to RIIPI, a D-C potential of $m^{th}$ order can be represented as a product of a multiplier $f_m$ and the $\operatorname{D-D_{IP}}$ potential of $\left(m-1\right)^{th}$ order, $\Pi_{\operatorname{D-D_{IP}}}^{m-1}$. Again, the problem is that an analytical $\Pi_{\operatorname{D-D_{IP}}}^m$ law can be derived only for even values of exponent $m$ leading to the well-defined D-C laws for $m=5,7,9..$. For even exponents $m=6,8,...$, D-C laws must be approximated in general.

One such approximation in the form of an infinite series is already discussed in Subsection \ref{inplanedisks}, see Eq.~\eqref{eqsingleser}. All the observations related to the $\operatorname{D-D_{IP}}$ law of order $m$ are valid for the D-C interactions of order $m+1$. For example, the D-C interaction potential for even $m$ can be represented analytically via the generalized hypergeometric function for the special case $R_x=R_y$, see Eq.~\eqref{eq:DDIPspec}.

The approximate P-C expressions for $m=4,6,8,...$ are derived in Subsection \ref{secPC}. Let us use these laws to derive approximate D-C laws with similar desired properties, cf.~Section \ref{secintro}.
By using RPCS1, the integral to be solved consists of the $arctanh$ and $arccos$ functions, i.e.
\begin{equation}
	\begin{aligned}
		\Pi_{\operatorname{D-C}}^m& \approx \tilde{\Pi}_{\operatorname{D-C_{app}}}^m= 2 \int _{R_x+q_2}^{R_x+q_2+2 R_y}  \tilde{\Pi}_{\operatorname{P-C_{app}}}^m \left[\textrm{arctanh}(t)\right] \arccos(\frac{t^2+d^2-R_y^2}{2td})t \dd{t}.
	\end{aligned}
\end{equation}
The square brackets in this expression emphasize that $ \tilde{\Pi}_{\operatorname{P-C_{app}}}^m $ is a function of $ arctanh (t) $. To the best of our knowledge, this integral cannot be solved analytically and we have to introduce additional assumptions. There are multiple options for an approximation and we present one that has a good balance between accuracy and efficiency.

We exclusively consider vdW attraction here ($m=6$). Let us assume that $R_x \ge R_y$ and introduce the following two approximations: (i) replace $t \, arccos (t) $  with its first order series expansion at $R_x+q_2$, and (ii) replace $arctanh(t)$ with its second order series expansion at $\infty$, see \hyperref[supsec]{Notebook 4}. The resulting expression can be integrated analytically, i.e.
\begin{equation}
	\label{eq:DCapp}
	\begin{aligned}
		\tilde{\Pi}_{\operatorname{D-C_{app}}}^6= \frac{\pi}{1920} \sqrt{\frac{d_3}{d}}\left(f_R+ \frac{20\sqrt{2 R_y}}{ R_x^2} f_T+ \frac{i \sqrt{2} R_y }{d_3 R_x^2 \sqrt{d_4 R_y}} f_E\right)
	\end{aligned}
\end{equation}
where the following new functions are introduced,
%
%
%
\begin{equation}
	\begin{aligned}
		f_R&\coloneqq 160\left(\frac{1}{q_2}  - \frac{1}{d_5}\right) -\frac{77}{\sqrt{d_1 d_2}} + \frac{11}{R_x} \left(\frac{\sqrt{d_1 d_2} \left(7d_4-22d_3\right)}{d_3 d_4} + 15\sqrt{\frac{d_1}{d_2}}\right) \\
		&+\frac{2 R_y}{R_x^2}\left[\frac{1512 \sqrt{d_1}}{\sqrt{d_2}}  + R_x \left(\frac{242 \sqrt{d_1}}{d_2^{3/2}}+\frac{180}{d_2} -  \frac{77 d_2}{d_1^3} + \frac{291}{\sqrt{d_1 d_2}} + \frac{160 \left(3q_2-2 R_x + 6 R_y\right)}{d_5^2}\right)\right]
	\end{aligned}
\end{equation}
%
%
%
%
%
\begin{equation}
	\begin{aligned}
		f_T&\coloneqq  \frac{9}{\sqrt{d_4}}  \left(2 q_2 + 3  R_x\right) \arctan{\sqrt{\tfrac{2 R_y}{d_4}}} - \frac{2 }{q_2^{3/2}} \left(9 q_2^2 + 12 q_2 R_x - 4 R_x^2 \right) \arctan{\sqrt{\tfrac{2 R_y}{q_2}}} 
	\end{aligned}
\end{equation}
\begin{equation}
	\begin{aligned}
		f_E&\coloneqq  \left(1512 q_2^2 + 4371 q_2 R_x +2936 R_x^2 \right) E(i \, \text{arcsinh} \sqrt{\tfrac{2 R_y}{d_3}},\tfrac{d_3}{d_4}) \\
		&- 11 R_x \left(96 q_2 +103 R_x\right) F(i \, \text{arcsinh} \sqrt{\tfrac{2 R_y}{d_3}},\tfrac{d_3}{d_4}).
	\end{aligned}
\end{equation}
Although \eqref{eq:DCapp} does not satisfy all of the desired properties from Section \ref{secintro}, the obtained expression is analytical and easy to evaluate in WM14. It is significantly more efficient than the infinite series solution in Eq.~\eqref{eqsingleser} or numerical integration. 

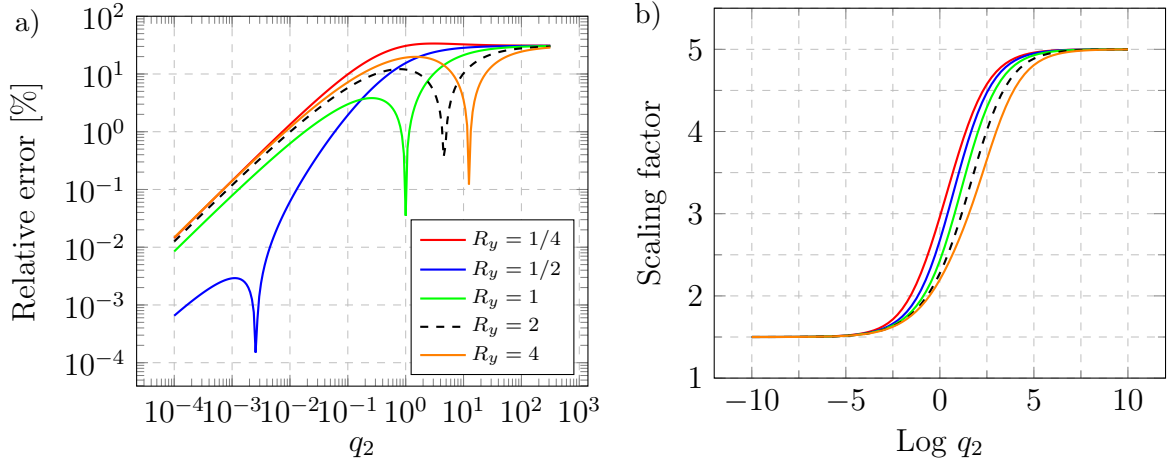
\begin{figure}[h!]
	\centering
	\begin{tikzpicture}
		\begin{loglogaxis}[
			xlabel = {$q_2$},
			ylabel = {Relative error $\left[\%\right]$},
			ylabel near ticks,
			legend pos=south east,
			legend cell align=left,
			legend style={font=\scriptsize,nodes={scale=0.95, transform shape}},
			width=0.47\textwidth,
			ymin = 0, ymax = 100,
			minor y tick num = 1,
			ytick distance=10^1,
			grid=major,
			clip=false];
			\node [text width=1em,anchor=north west] at (rel axis cs: -0.3,1.1){\subcaption{\label{fig:a}}};
			\addplot[red,thick] table [col sep=comma] {data/dataCylCyl6Error1.csv};
			\addplot[blue,thick] table [col sep=comma] {data/dataCylCyl6Error2.csv};
			\addplot[green,thick] table [col sep=comma] {data/dataCylCyl6Error3.csv};
			\addplot[black,dashed,thick] table [col sep=comma] {data/dataCylCyl6Error4.csv};
			\addplot[orange,thick] table [col sep=comma] {data/dataCylCyl6Error5.csv};
			\legend{$R_y=1/4$,$R_y=1/2$,$R_y=1$,$R_y=2$,$R_y=4$} 
		\end{loglogaxis}
	\end{tikzpicture}
	\begin{tikzpicture}
		\begin{axis}[
			xlabel = {Log $q_2$},
			ylabel = {Scaling factor},
			ylabel near ticks,
			legend pos=south east,
			width=0.47\textwidth,
			ymin = 1, ymax = 5.5,
			minor y tick num = 1,
			minor x tick num = 1,
			ytick distance=1,
			grid=both,
			clip=false]
			\node [text width=1em,anchor=north west] at (rel axis cs: -0.205,1.1){\subcaption{\label{fig:b}}};
			\addplot[red,thick] table [col sep=comma] {data/dataCylCyl6Scaling1.csv};
			\addplot[blue,thick] table [col sep=comma] {data/dataCylCyl6Scaling2.csv};
			\addplot[green,thick] table [col sep=comma] {data/dataCylCyl6Scaling3.csv};
			\addplot[black,dashed,thick] table [col sep=comma] {data/dataCylCyl6Scaling4.csv};
			\addplot[orange,thick] table [col sep=comma] {data/dataCylCyl6Scaling5.csv};
		\end{axis}
	\end{tikzpicture}
	\caption{Approximate D-C$_{\text{app}}$ law for $m=6$. The radius of the cylinder $X$ is set as 1, while the radius of the disk $Y$ is varied. a) Relative error. b) Scaling factor. }
	\label{fig:DCapp}
\end{figure}

Fig.~\ref{fig:DCapp} shows the error and the scaling of this new D-C$_{\text{app}}$ vdW law. 
The error behaves similarly to the error of the P-C$_{\text{app}}$ law: it is small for small separations and bounded for large separations, with a maximum around 30 \%, see Fig.~\ref{fig:DCapp}a. The error is not a monotonic function of the gap $q_2$ but shows an increase in accuracy at localized regions. We can observe that the expression is not symmetric, e.g. the errors for $R_y=1/2$ and $R_y=2$ are not the same, although they should be for $R_x=1$. The asymptotic scaling is correct: $3/2$ for small and $5$ for large separations. The ratio between radii affects both the accuracy and the scaling. All in all, the new D-C$_{\text{app}}$ law is not ideal but provides a reasonable balance between accuracy and efficiency.


\subsection{Disk-plate interaction for arbitrary $m$}
\label{diskplatesub}

\begin{figure}[h!]
	\begin{center}
		\includegraphics[width=0.45\linewidth]{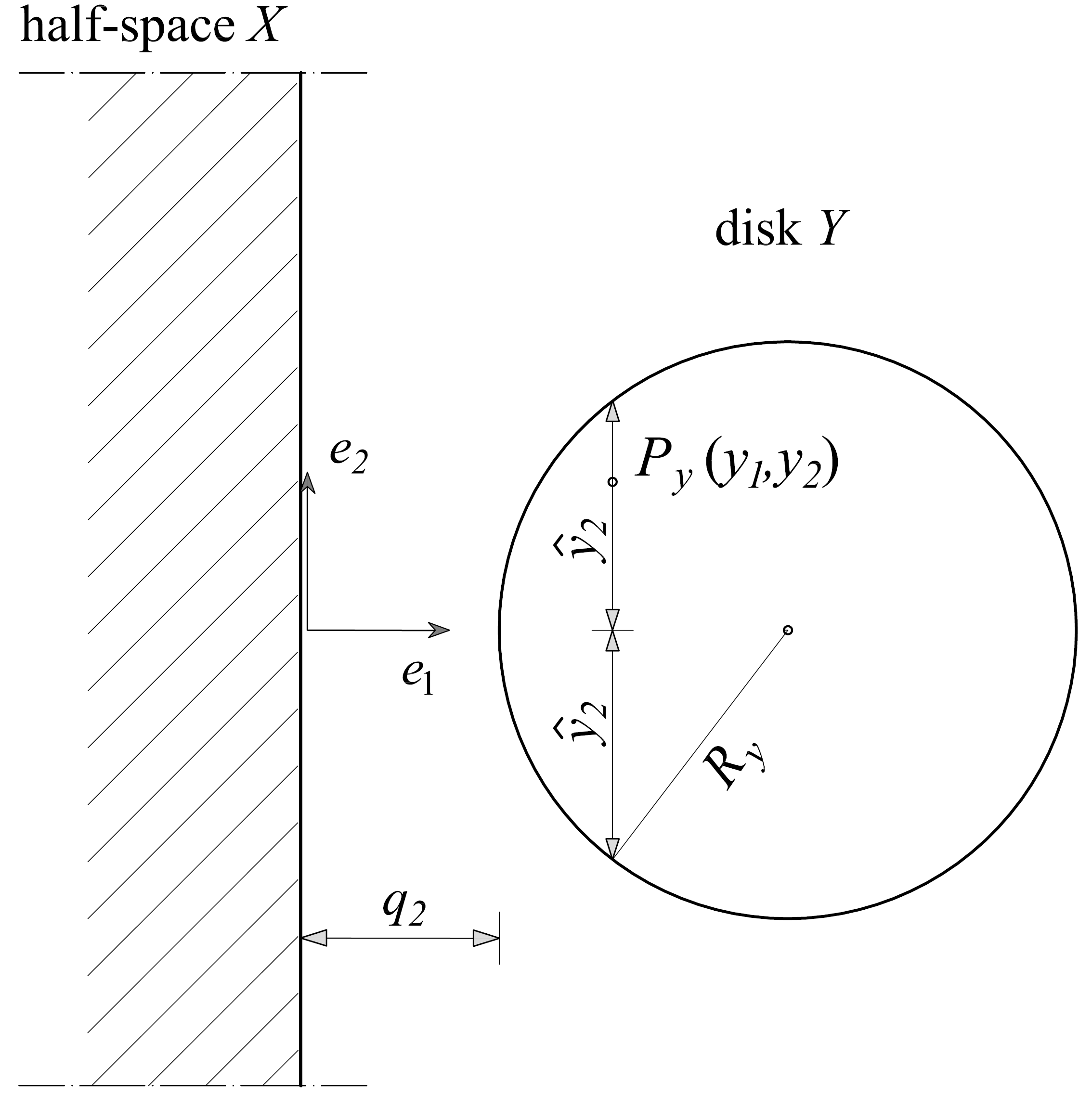}
		\caption{Interaction between a half-space $X$ and a disk $Y$ in perpendicular orientation.}
		\label{fig:DHS}
	\end{center}
\end{figure}

Let us consider an interaction between a half-space $X$ and a disk $Y$ in perpendicular orientation as in Fig.~\ref{fig:DHS}.
The origin of the Cartesian coordinate system is at the boundary of the infinite half-space $X$. Then, we utilize the P-HS law for general $m$ \eqref{eqP-HS} and replace the gap $q_2$ with the coordinate $y_1$. In this way, the integration limits are $y_1 \in [q_2,q_2+2R_y]$ and $y_2 \in [-\hat{y}_2,\hat{y}_2]$, where $\hat{y_2}=\sqrt{R_y^2-(R_y+q_2-y_1)^2}$. With this setting, we can find the general analytical solution for the D-HS interaction, i.e.
\begin{equation}
	\begin{aligned}
		\Pi_{\operatorname{D-HS}}^m&=  \int_{A_y} \Pi_{\operatorname{P-HS}}^m \dd{A_y} =  \int_{q_2}^{q_2+2R_y} \int_{-\sqrt{R_y^2-(R_y+q_2-y_1)^2}}^{\sqrt{R_y^2-(R_y+q_2-y_1)^2}} \frac{2 \pi }{\left(m-3\right) \left(m-2\right) y_1^{m-3}} \dd{y_2}\dd{y_1} \\
		&= \frac{\pi^2 R_y^2}{2\left(m-3\right)\left(m-2\right)q_2^m} \left[4 q_2^3 \, _2F_1 \left(\frac{3}{2},m;3;-\frac{2R_y}{q_2}\right) +12 q_2^2 R_y  \, _2F_1 \left(\frac{5}{2},m;4;-\frac{2R_y}{q_2}\right)  \right. \\
		&\left. + 15 q_2 R_y^2 \, _2F_1 \left(\frac{7}{2},m;5;-\frac{2R_y}{q_2}\right) + 7 R_y^3 \, _2F_1 \left(\frac{9}{2},m;6;-\frac{2R_y}{q_2}\right) \right] \; \text{for} \; m>\frac{9}{2}.
	\end{aligned}
	\label{eq:PDHS}
\end{equation}
Due to IUI, a D-HS interaction is also an interaction between a cylinder of unit length and an infinite half-space. The asymptotic scaling is $m - 4.5$ for small, and $m - 3$ for large separations. For integer values of $m$, this general expression simplifies to rational functions. For example, the vdW potential is obtained by inserting $m=6$ in \eqref{eq:PDHS}, which gives
\begin{equation}
	\begin{aligned}
		\Pi_{\operatorname{D-HS}}^6&= \frac{\pi^2 R_y^2}{6 \left[q_2\left(q_2+2 R_y\right)\right]^{3/2}}.
	\end{aligned}
\end{equation}
For small separation distances, the vdW D-HS interaction potential further simplifies to
\begin{equation}
	\begin{aligned}
		\Pi_{\operatorname{D-HS}}^6&\approx \frac{\pi^2 \sqrt{R_y}}{12 \sqrt{2} q_2^{3/2}}\; \text{for} \; q_2 \ll R_y
	\end{aligned}
\end{equation}
which is a well-known expression \cite{2011israelachvili}.

\begin{figure}[h]
	\centering
	\begin{tikzpicture}
		\begin{axis}[
			xlabel = {Log $q_2$},
			ylabel = {Scaling factor},
			ylabel near ticks,
			legend pos=outer north east,
			legend cell align=left,
			legend style={font=\tiny},
			width=0.39\textwidth,
			height=0.4\textwidth,
		minor y tick num = 1,
		minor x tick num = 1,
			xtick distance=5,
				ytick distance=2,
			grid=both,
			clip=false];
			\node [text width=1em,anchor=north west] at (rel axis cs: -0.31,1.1){\subcaption{\label{fig:a}}};
			\addplot[red,thick] table [col sep=comma] {data/dataDPTmScaling1.csv};
			\addplot[red,dashed,thick] table [col sep=comma] {data/dataDPTmScaling2.csv};
			\addplot[blue,thick] table [col sep=comma] {data/dataDPTmScaling3.csv};
			\addplot[blue,dashed,thick] table [col sep=comma] {data/dataDPTmScaling4.csv};
			\addplot[orange,thick] table [col sep=comma] {data/dataDPTmScaling5.csv};
			\addplot[orange,dashed,thick] table [col sep=comma] {data/dataDPTmScaling6.csv};
			\addplot[green,thick] table [col sep=comma] {data/dataDPTmScaling7.csv};
			\addplot[green,dashed,thick] table [col sep=comma] {data/dataDPTmScaling8.csv};
			\legend{$m=5$,$m=6$,$m=7$,$m=8$,$m=9$,$m=10$,$m=11$,$m=12
			$} 
		\end{axis}
	\end{tikzpicture}
	\begin{tikzpicture}
		\begin{axis}[
			xlabel = {Log $q_2$},
			ylabel = {Scaling factor},
			ylabel near ticks,
			legend pos=south east,
			legend style={font=\tiny},
			legend cell align=left,
			width=0.46\textwidth,
			xtick distance=5,
			minor y tick num = 1,
			minor x tick num = 1,
			grid=both,
			clip=false]
			\node [text width=1em,anchor=north west] at (rel axis cs: -0.21,1.1){\subcaption{\label{fig:b}}};
			\addplot[red,thick] table [col sep=comma] {data/dataDPT6Scaling1.csv};
			\addplot[blue,thick] table [col sep=comma] {data/dataDPT6Scaling2.csv};
			\addplot[orange,thick] table [col sep=comma] {data/dataDPT6Scaling3.csv};
			\addplot[green,thick] table [col sep=comma] {data/dataDPT6Scaling4.csv};
			\addplot[black,thick] table [col sep=comma] {data/dataDPT6Scaling5.csv};
			\legend{$b_y=10^{-1}$,$b_y=10^{0}$,$b_y=10^{1}$,$b_y=10^{2}$,$b_y=10^{3}$}
		\end{axis}
	\end{tikzpicture}
	\caption{General D-PT law, $R_x=1$. a) Scaling factor for different exponents $m$ and $b_y=1$. b) vdW scaling factor for different values of $b_y$. }
	\label{fig:DPT}
\end{figure}
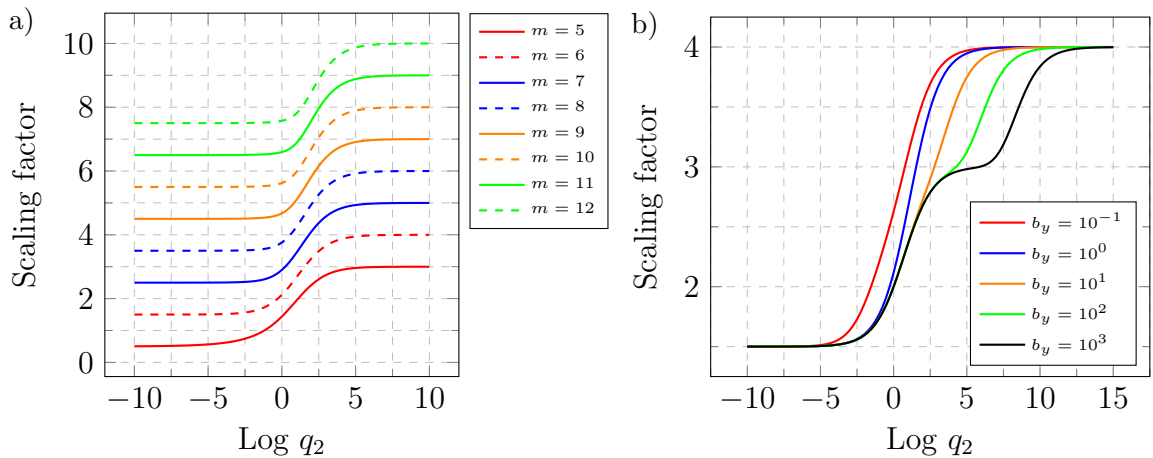

From the general D-HS law \eqref{eq:PDHS}, it is straightforward to find the general D-PT law as a special case of RIHBI. If the thickness of the plate is $2 b_y$, the general D-PT law is
\begin{equation}
	\begin{aligned}
		\Pi_{\operatorname{D-PT}}^m (q_2,b_y)&=  \Pi_{\operatorname{D-HS}}^m \left(q_2\right) - \Pi_{\operatorname{D-HS}}^m \left(q_2+2b_y\right)\; \text{for} \; m>\frac{9}{2}.
	\end{aligned}
\end{equation}
The scaling factor of the D-PT law is displayed in Fig.~\ref{fig:DPT}. 
For small separations, the potential scales with $m-4.5$, and for large with $m-2$, cf. Fig.~\ref{fig:DPT}a. As expected, the asymptotic scaling factor at small separations is the same for the interaction of a disk with an infinite half-space and with a plate. Furthermore, let us observe the scaling behavior for $m=6$ and vary the thickness of the plate, see Fig.~\ref{fig:DPT}b. We can observe how, for large values of thickness and separation, the scaling factor first approaches $m-3=3$, which corresponds to the D-HS law, and then converges to the correct D-PT value of $m-2=4$.

%
%
%
%

\section{Rectangular cross sections}
\label{rectsection}

For an interaction between two rectangular sections, we consider the special case where the centers (or their graphical projections) of both sections lie in the same plane, and the sections are parallel to each other, see Fig.~\ref{fig:intro1}d. The dimensions of the sections are denoted by $2b_x \times 2h_x$ and $2b_y \times 2h_y$.

\subsection{In-plane rectangle-rectangle interaction for arbitrary $m$}

\begin{figure}[h!]
	\begin{center}
		\includegraphics[width=0.8\linewidth]{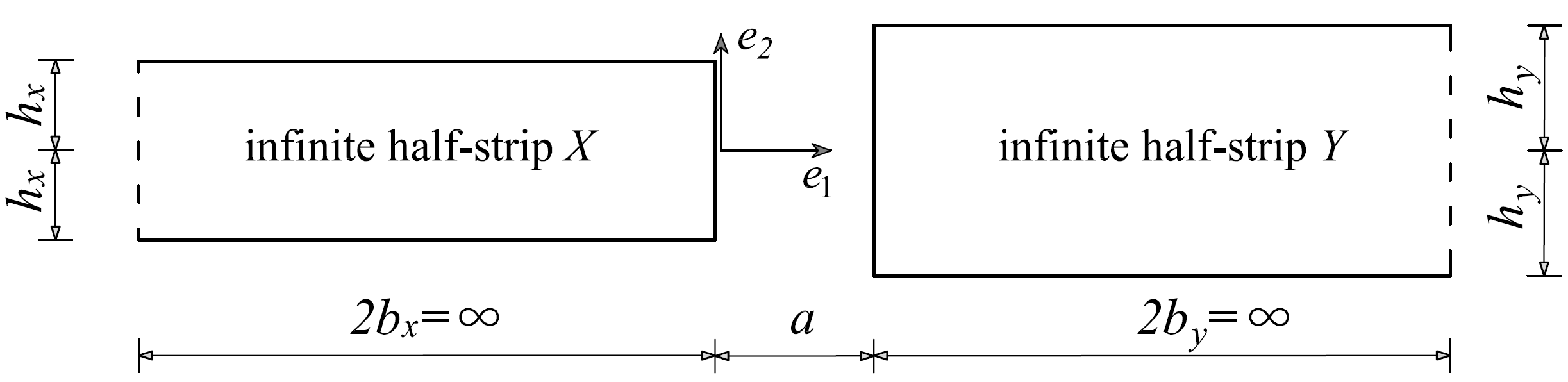}
		\caption{Interaction between two infinite in-plane half-strips, HST-HST$_\text{IP}$}
		\label{fig:HRHR}
	\end{center}
\end{figure}

To find the R-R$_{\text{IP}}$ laws, we can employ the Cartesian coordinate system in Fig.~\ref{fig:intro1}d so that the integral to solve is
\begin{equation}
	\begin{aligned}
		\Pi_{\operatorname{R-R_{IP}}}^m&=\int_{q_2+b_x}^{q_2+b_x + 2 b_y} \int_{-h_y}^{h_y} \int_{-b_x}^{b_x} \int_{-h_x}^{h_x} p^{-m} \dd{x_2}\dd{x_1} \dd{y_2}\dd{y_1}.
	\end{aligned}
	\label{eq:RRIP}
\end{equation}
We can solve this integral directly for every specific integer value $m=4,5,6..$. However, there is a more general approach. Let us define an infinite half-strip as one half of a rectangle that has one side of infinite length, see Fig.~\ref{fig:HRHR}. 
If we first derive an in-plane infinite half-strip-infinite half-strip ($\operatorname{HST-HST_{IP}}$) law for general integer $m$, and apply RIHBI, we can obtain a general R-R$_{\text{IP}}$ law.

Therefore, let us find an interaction potential between two infinite in-plane half-strips, $\Pi_{\operatorname{HST-HST_{IP}}}^m$. A scheme of this interaction and the coordinate system are depicted in Fig.~\ref{fig:HRHR}. To avoid ambiguity, we designate the gap between these two infinite half-strips with $a$, i.e.
\begin{equation}
	\begin{aligned}
		\Pi_{\operatorname{HST-HST_{IP}}}^m (a)&=\int_{a}^{\infty} \int_{-h_y}^{h_y} \int_{-\infty}^{0} \int_{-h_x}^{h_x} p^{-m}  \dd{x_2}\dd{x_1} \dd{y_2}\dd{y_1}.
	\end{aligned}
	\label{eq:RRIP1}
\end{equation}
We can not solve this integral directly for general integer $m$ in WM14. But, we can find the analytical solutions for two distinct cases, odd and even $m$. It turns out that both of these expressions are the same and valid for arbitrary $m \ge 4$, see \hyperref[supsec]{Notebook 5}. So, the general $\operatorname{HST-HST_{IP}}$ law is
\begin{equation}
	\begin{aligned}
		\tilde{\Pi}_{\operatorname{HST-HST_{IP}}}^m (a)&=f(a,1) +f(a,-1)+g(a,1) +g(a,-1)+u(a,1) +u(a,-1)\; \text{for} \; m \ge 4,
	\end{aligned}
	\label{eq:RRIP2}
\end{equation}
where
\begin{equation}
	\begin{aligned}
		f(a,s) &\coloneqq \frac{2\, s \, k(s)^2 \, a^{2-m}}{(m-3)(m-2)^2(m-1)}  \,_2F_1 \left(\frac{m-2}{2},\frac{m-1}{2};\frac{m+1}{2};-\frac{k(s)^2}{a^2}\right) \\
		g(a,s) &\coloneqq \frac{2\, s \, k(s)^2 \, a^{2-m}}{(m-2)^2}  \,_2F_1 \left(\frac{1}{2},\frac{m-2}{2};\frac{3}{2};-\frac{k(s)^2}{a^2}\right) \\
		u(a,s) &\coloneqq \frac{2\, s  \left(k(s)^2 + a^{2}\right)^{2-m/2}}{(m-4)(m-3)(m-2)}  \\
		k\left(s\right) &\coloneqq h_x+sh_y.
	\end{aligned}
	\label{eq:RRIP3}
\end{equation}
This expression is accurate and can be evaluated efficiently for every $m >4$. However, for $m=4$ it has a singularity, see function $u(a,s)$ in \eqref{eq:RRIP3}. This singularity can be fixed by finding the limit of $\tilde{\Pi}_{\operatorname{HST-HST_{IP}}}^m (a)$ for $m \rightarrow 4$, i.e.
\begin{equation}
	\begin{aligned}
		\tilde{\Pi}_{\operatorname{HST-HST_{IP}}}^4 (a) =\lim\limits_{m \rightarrow 4} \tilde{\Pi}_{\operatorname{HST-HST_{IP}}}^m (a) &= \frac{\left(\frac{a^2}{k(-1)}-k(-1)\right)\arctan(\frac{k(-1)}{a})}{2a} \\
		&+\frac{\left(k(1)-a^2\right)\arctan(\frac{k(1)}{a})}{2a k(1)} 
		+ \frac{1}{2}\ln(\frac{k(-1) + a^2}{k(1)+a^2}).
	\end{aligned}
	\label{eq:RRIP4}
\end{equation}
This allows us to write the general $\operatorname{HST-HST_{IP}}$ law as a piecewise function,
\begin{equation}
	\begin{aligned}
		\Pi_{\operatorname{HST-HST_{IP}}}^m (a) &=
		\begin{cases} 
			\Pi_{\operatorname{HST-HST_{IP}}}^4 (a) & m=4 \\
			\tilde{\Pi}_{\operatorname{HST-HST_{IP}}}^m (a) & m>4
		\end{cases}.
	\end{aligned}
	\label{eq:RRIP5}
\end{equation}
There is another issue with the above expression, a singularity for $h_x = h_y$. However, the limit for $h_x \rightarrow h_y$ is well-defined and it can be used in special cases, i.e.
\begin{equation}
	\begin{aligned}
		\hat{\Pi}_{\operatorname{HST-HST_{IP}}}^m (a) &= \lim_{h_x \rightarrow h_y} \Pi_{\operatorname{HST-HST_{IP}}}^m (a) =
		\begin{cases} 
			R_{h_x=h_y}^4 & m=4 \\
			R_{h_x=h_y}^m & m>4
		\end{cases}, \\
	R_{h_x=h_y}^4 &\coloneqq \frac{1}{2} \left(1+\left(\frac{2 h_y}{a} - \frac{a}{2 h_y}\right) \arctan(\frac{2h_y}{a})+\ln(\frac{a^2}{4hy^2+a^2})\right), \\
	R_{h_x=h_y}^m &\coloneqq\frac{2\left(\left(4h_y^2+a^2\right)^{2-\frac{m}{2}}-a^{4-m}\right)}{(m-4)(m-3)(m-2)} \\
	&+  \frac{8 h_y^2 a^{2-m}}{(m-2)^2} \left(_2F_1 \left(\frac{1}{2},\frac{m-2}{2};\frac{3}{2};-\frac{4h_y^2}{a^2}\right)+\frac{ _2F_1 \left(\frac{m-2}{2},\frac{m-1}{2};\frac{m+1}{2};-\frac{4 h_y^2}{a^2}\right)}{(m-3)(m-1)}\right).
	\end{aligned}
	\label{eq:RRIP6}
\end{equation}
With such well-defined $\operatorname{HST-HST_{IP}}$ law as a function of gap $a$, we can apply RIHBI and obtain the general R-R$_{\text{IP}}$ law, i.e.
\begin{equation}
	\begin{aligned}
		\Pi_{\operatorname{R-R_{IP}}}^m (q_2,b_x,b_y)&= \Pi_{\operatorname{HST-HST_{IP}}}^m \left(q_2\right)-\Pi_{\operatorname{HST-HST_{IP}}}^m \left(2 b_x+q_2\right) \\
		&-\Pi_{\operatorname{HST-HST_{IP}}}^m \left(2b_y+q_2\right)+\Pi_{\operatorname{HST-HST_{IP}}}^m \left(2b_x+2b_y+q_2\right) \; \text{for} \; m\ge 4.
	\end{aligned}
	\label{eq:RRIP7}
\end{equation}
For the special case $h_x=h_y$, we need to use expression \eqref{eq:RRIP6}. 

The obtained expressions 
satisfy all the requirements stated in Section \ref{secintro}. They reduce to simple rational functions for integer values of $m$. For example, we can compare our result with the special case from the literature where equal rectangles ($h_x=h_y$, $b_x=b_y$) with $m=6$ are considered \cite{1960derocco}. Our general expression returns
\begin{equation}
	\begin{aligned}
		\hat{\Pi}_{\operatorname{R-R_{IP}}}^6&= \frac{1}{48} \left(\frac{2}{(2 b_y +q_2)^2} - \frac{1}{(4b_y+q_2)^2} - \frac{1}{q_2^2}\right) + \frac{1}{128 h_y^3} \left[ \left(\frac{16 h_y^4}{q_2^3}-q_2\right) \arctan(\frac{2 h_y}{q_2}) \right. \\
		&+ \left(4 b_y + 2 q_2 - \frac{32 h_y^4}{(2b_y+q_2)^3}\right) \arctan(\frac{2 h_y}{2 b_y+q_2}) \\
		&\left. - \left(4 b_y + q_2 - \frac{16 h_y^4}{(4b_y+q_2)^3}\right) \arctan(\frac{2 h_y}{4 b_y+q_2})\right],
	\end{aligned}
	\label{eq:RRIP8}
\end{equation}
which  is the same as in \cite{1960derocco}, just in a different form. This confirms that our general R-R$_{\text{IP}}$ law correctly models the interaction between rectangles with arbitrary dimensions. 

\begin{figure}[h]
	\centering
	\begin{tikzpicture}
		\begin{axis}[
			xlabel = {Log $q_2$},
			ylabel = {Scaling factor},
			ylabel near ticks,
			legend pos=outer north east,
			legend cell align=left,
			legend style={font=\tiny},
			width=0.4\textwidth,
			height=0.4\textwidth,
			minor y tick num = 1,
			minor x tick num = 1,
			xtick distance=5,
			ytick distance=2,
			grid=both,
			clip=false];
			\node [text width=1em,anchor=north west] at (rel axis cs: -0.30,1.1){\subcaption{\label{fig:a}}};
			\addplot[red,thick] table [col sep=comma] {data/dataRRPscaling1.csv};
			\addplot[red,dashed,thick] table [col sep=comma] {data/dataRRPscaling2.csv};
			\addplot[blue,thick] table [col sep=comma] {data/dataRRPscaling3.csv};
			\addplot[blue,dashed,thick] table [col sep=comma] {data/dataRRPscaling4.csv};
			\addplot[orange,thick] table [col sep=comma] {data/dataRRPscaling5.csv};
			\addplot[orange,dashed,thick] table [col sep=comma] {data/dataRRPscaling6.csv};
			\addplot[green,thick] table [col sep=comma] {data/dataRRPscaling7.csv};
			\addplot[green,dashed,thick] table [col sep=comma] {data/dataRRPscaling8.csv};
			\legend{$m=5$,$m=6$,$m=7$,$m=8$,$m=9$,$m=10$,$m=11$,$m=12
				$} 
		\end{axis}
	\end{tikzpicture}
	\begin{tikzpicture}
		\begin{axis}[
			xlabel = {Log $q_2$},
			ylabel = {Scaling factor},
			ylabel near ticks,
			legend pos=north west,
			legend style={font=\tiny,nodes={scale=0.92, transform shape}},
			legend cell align=left,
			width=0.46\textwidth,
			minor y tick num = 1,
			minor x tick num = 1,
			xtick distance=5,
			ytick distance=1,
			grid=both,
			clip=false]
			\node [text width=1em,anchor=north west] at (rel axis cs: -0.21,1.1){\subcaption{\label{fig:b}}};
			\addplot[red,dashed,thick] table [col sep=comma] {data/dataRRP6scaling1.csv};
			\addplot[blue,thick] table [col sep=comma] {data/dataRRP6scaling2.csv};
			\addplot[green, dashed,thick] table [col sep=comma] {data/dataRRP6scaling3.csv};
			\addplot[black,thick] table [col sep=comma] {data/dataRRP6scaling4.csv};
			\addplot[orange, dashed,thick] table [col sep=comma] {data/dataRRP6scaling5.csv};
			\addplot[green,thick] table [col sep=comma] {data/dataRRP6scaling6.csv};
			\addplot[blue,dashed,thick] table [col sep=comma] {data/dataRRP6scaling7.csv};
			\legend{$b_y=10^{-3}$,$b_y=10^{-2}$,$b_y=10^{-1}$,$b_y=10^{0}$,$b_y=10^{1}$,$b_y=10^{2}$,$b_y=10^{3}$}
		\end{axis}
	\end{tikzpicture}
	\caption{Exact R-RP laws. a) Scaling factor for different values of $m$. Dimensions are fixed: $h_x=1$, $h_y=1/2$, $b_x=1$, and $b_y=2$. b) Scaling factor for $m=6$. Dimensions $h_x=1$, $h_y=1/2$, and $b_x=1$ are fixed, while $b_y$ is varied.}
	\label{fig:RRP}
\end{figure}
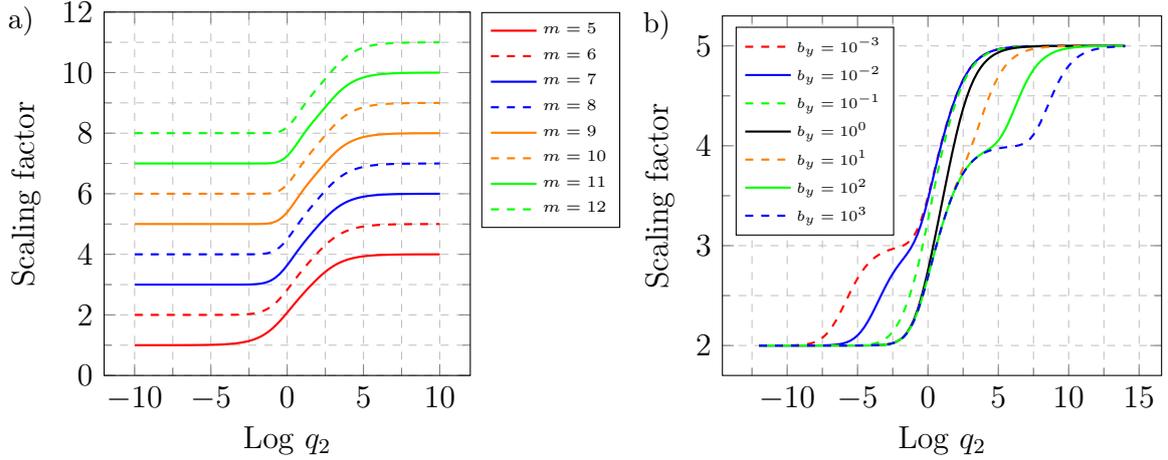

From the general R-R$_{\text{IP}}$ law for $m \ge 4$, we can simply obtain R-RP laws for $m \ge 5$ by using RIIPI from Eq.~\eqref{eq1rule1}, i.e.
\begin{equation}
	\begin{aligned}
		\Pi_{\operatorname{R-RP}}^m&=f_m \Pi_{\operatorname{R-R_{IP}}}^{m-1}. 
	\end{aligned}
	\label{eq:RRIP9}
\end{equation}
The scaling behavior of the R-RP laws for $m=5,6,...12$ is plotted in Fig.~\ref{fig:RRP}a. 
We can observe that, for small separations, the asymptotic scaling factor is $m-4$, while for large it is $m-1$, because then the rectangle and the prism resemble a point and a line. The transition region between small and large separations is dependent on $m$ and cannot be obtained by simply shifting the function for some other exponent.

As a special case, let us consider vdW attraction, i.e. $m=6$. After some simplification, the R-RP law can be represented as
\begin{equation}
	\begin{aligned}
		\Pi_{\operatorname{R-RP}}^6&= \xi \left(q_2\right)-\xi\left(2 b_x+q_2\right)-\xi\left(2b_y+q_2\right)+\xi\left(2b_x+2b_y+q_2\right) \\
		\xi \left(a\right)&\coloneqq\frac{\pi}{12} \left(\frac{\left[a^2+\left(h_x+h_y\right)^2\right]^{3/2}}{a^2 \left(h_x+h_y\right)^2}- \frac{\left[a^2+\left(h_x-h_y\right)^2\right]^{3/2}}{a^2 \left(h_x-h_y\right)^2} \right).
	\end{aligned}
	\label{RRPvdW}
\end{equation}
The function $\xi$ is not just equal to $f_6 \Pi_{\operatorname{R_{\infty}-R_{\infty IP}}}^5$, because some terms cancel out when we apply RIHBI and the final expression simplifies. The singularity for $h_x = h_y$ can be addressed in two ways: (i) by using Eq.~\eqref{eq:RRIP6} or (ii) by transforming the expression. Let us first use Eq.~\eqref{eq:RRIP6} and transform the expression in the next subsection, i.e.
\begin{equation}
	\begin{aligned}
		\hat{\Pi}_{\operatorname{R-RP}}^6&= \hat{\xi} \left(q_2\right)-\hat{\xi}\left(2 b_x+q_2\right)-\hat{\xi}\left(2b_y+q_2\right)+\hat{\xi}\left(2b_x+2b_y+q_2\right) \\
		\hat{\xi} \left(a\right)&\coloneqq\frac{\pi \left( (4 h_y^2 + a^2)^{3/2}-6 h_y^2 a\right)}{48 h_y^2 a^2}.
	\end{aligned}
	\label{RRPvdWspec}
\end{equation}
The scaling of the vdW case for different values of $b_y$ is given in Fig.~\ref{fig:RRP}b. For small separations and small values of $b_y$, the scaling factor first approaches $m-3=3$, which corresponds to the interaction between a rectangle and a prism with infinitely small thickness $b_y$, i.e.~a strip. Then it approaches 2. Analogously, for large separations and large $b_y$, the scaling first approaches value $m-2=4$, which corresponds to the interaction of a rectangle and a prism with an infinite thickness $b_y$, i.e.~an infinite half-plate. Then it approaches 5.


\subsection{Exact rectangle-rectangle law for vdW attraction}

In this subsection, we will consider a general (with offset $q_1$) rectangle-rectangle interaction that returns several special cases. We will focus only on vdW attraction, i.e.~$m=6$. The integral to solve is:
\begin{equation}
	\begin{aligned}
		\Pi_{\operatorname{R-R}}^6&=\int_{q_2+b_x}^{q_2+b_x + 2 b_y} \int_{-h_y}^{h_y} \int_{-b_x}^{b_x} \int_{-h_x}^{h_x} \left(p^2+q_1^2\right)^{-3}  \dd{x_2}\dd{x_1} \dd{y_2}\dd{y_1}.
	\end{aligned}
	\label{rec1}
\end{equation}
As in the previous subsection, we can solve this integral directly, but a more elegant approach is to apply RIHBI. Therefore, let us first find an interaction law between two infinite half-strips with an offset. We use the setup as in Fig.~\ref{fig:HRHR}, i.e.
\begin{equation}
	\begin{aligned}
		\Pi_{\operatorname{HST-HST}}^6 (a)&=\int_{a}^{\infty} \int_{-h_y}^{h_y} \int_{-\infty}^{0} \int_{-h_x}^{h_x} \left(p^2+q_1^2\right)^{-3}  \dd{x_2}\dd{x_1} \dd{y_2}\dd{y_1} = e\left(a\right) +j\left(a\right),
	\end{aligned}
	\label{rec2}
\end{equation}
where
\begin{equation}
	\begin{aligned}
		e\left(a\right) &\coloneqq \frac{\pi a}{8 q_1^4}\left[\frac{q_1^2}{w\left(1\right)}-\frac{q_1^2}{w\left(-1\right)}+2w\left(-1\right)-2w\left(1\right)\right]\\
		j\left(a\right) &\coloneqq \frac{1}{8 q_1^4}\left[v\left(a,1\right) + v\left(a,-1\right) +l\left(a,1\right) - l\left(a,-1\right)\right] \\
		v\left(a,s\right) &\coloneqq \frac{2 s a \left[2 k^2\left(s\right) + q_1^2\right] \arctan\frac{a}{\sqrt{k^2\left(s\right)+q_1^2}}}{\sqrt{k^2\left(s\right) + q_1^2}} \\
		l\left(a,s\right) &\coloneqq \frac{2 k\left(s\right) \left(q_1^2+2a^2\right) \arctan\frac{k\left(s\right)}{\sqrt{q_1^2+a^2}}}{\sqrt{q_1^2 + a^2}} \\
		w\left(s\right) &\coloneqq \sqrt{k\left(s\right)^2+q_1^2}.
	\end{aligned}
	\label{eq:RRhelp}
\end{equation}
This law consists of two parts, one that is linear w.r.t.~the gap $a$ -- $e(a)$. Due to this, the function $e(a)$ cancels out when we apply RIHBI to the HST-HST law and the final elegant representation of the R-R law for $m=6$ is
%
\begin{equation}
	\begin{aligned}
		\Pi_{\operatorname{R-R}}^6&= j\left(q_2\right)-j\left(2 b_x+q_2\right)-j\left(2b_y+q_2\right)+j\left(2b_x+2b_y+q_2\right).
	\end{aligned}
	\label{eq:RR6}
\end{equation}
The obtained expression is both accurate and efficient but has a numerical issue for $q_1\approx 0$ which requires an arbitrary precision arithmetic. This issue can be addressed by introducing a simple piecewise function that, for example, uses machine precision and equation \eqref{eq:RR6} for, approximately, $\abs{q_1}>10^{-3}$, and an arbitrary precision or $\Pi_{\operatorname{R-R_{IP}}}^6$ otherwise. 

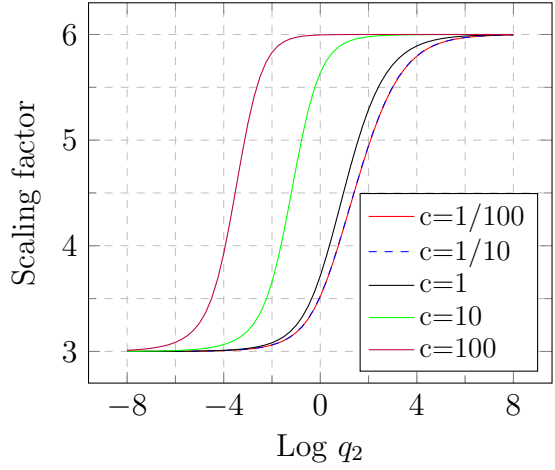
\begin{figure}[h]
	\centering
	\begin{tikzpicture}
		\begin{axis}[
			xlabel = {Log $q_2$},
			ylabel = {Scaling factor},
			legend pos=south east,
			legend cell align=left,
			ylabel near ticks,
			legend style={nodes={scale=0.9, transform shape}},
			width=0.45\textwidth,
			minor y tick num = 1,
			minor x tick num = 1,
			xtick distance=4,
			ytick distance=1,
			grid=both,
			width=0.48\textwidth,
			clip=false];
			\addplot[red] table [col sep=comma] {data/dataRR6scaling1.csv};
			\addplot[blue,dashed] table [col sep=comma] {data/dataRR6scaling2.csv};
			\addplot[black] table [col sep=comma] {data/dataRR6scaling3.csv};
			\addplot[green] table [col sep=comma] {data/dataRR6scaling4.csv};
			\addplot[purple] table [col sep=comma] {data/dataRR6scaling5.csv};
			\legend{c=1/100,c=1/10,c=1,c=10,c=100} 
		\end{axis}
	\end{tikzpicture}
	%
	\caption{Scaling of the R-R law for $m=6$ and different ratios $c=q_1/q_2$ (for $h_x=h_y=b_x=b_y=1$). 
	}
	\label{fig:RR6scaling}
\end{figure}

The scaling factor of the R-R law is presented in Fig.~\ref{fig:RR6scaling} for different values of ratio $c=q_1/q_2$. 
We can observe that the asymptotic scaling factor for small separations is $m-3=3$, and $m-0=6$ for large separations, as expected. The ratio $c$ has a significant influence on the scaling factor in the transition range from small to large separations. As for the D-D law, cf.~Fig.~\ref{fig:LJpoten3}a, it turns out that this transition region is invariant to the ratio $c$ when $c<1/10$.

By integrating $\Pi_{\operatorname{R-R}}^6$ w.r.t.~$q_1$ from $-L_y$ to $L_y$ we obtain an interaction law between a rectangle and a rectangular prism of a finite length $2L_y$. The same result follows if we integrate function $j\left(a\right)$ in \eqref{eq:RRhelp} and apply RIHBI, cf. \hyperref[supsec]{Notebook 6}. The final result is
\begin{equation}
	\begin{aligned}
		\Pi_{\operatorname{R-RP_{L_y}}}^6&= \int_{-L_y}^{L_y} \Pi_{\operatorname{R-R}}^6\dd{q_1} \\
		&= J\left(q_2\right)-J\left(2 b_x+q_2\right)-J\left(2b_y+q_2\right)+J\left(2b_x+2b_y+q_2\right),
	\end{aligned}
	\label{eq:RRPLy}
\end{equation}
where
\begin{equation}
	\begin{aligned}
		J\left(a\right) &\coloneqq F\left(a,1\right) + F\left(a,-1\right) +G\left(a,1\right) + G\left(a,-1\right) +P\left(a,1\right) + P\left(a,-1\right) \\
		F\left(a,s\right) &\coloneqq - \frac{s a \left[2 k^2\left(s\right) -L_y^2\right] \sqrt{k^2\left(s\right) + L_y^2} \arctan\frac{a}{\sqrt{k^2\left(s\right)+L_y^2}}}{6 k^2\left(s\right) L_y^3} \\
		G\left(a,s\right) &\coloneqq \frac{s \left[k^2\left(s\right) +a^2\right]^{3/2}  \arctan\frac{L_y}{\sqrt{k^2\left(s\right)+a^2}}}{6 k^2\left(s\right) a^2} \\
		P\left(a,s\right) &\coloneqq \frac{s k\left(s\right) \left(L_y^2-2a^2\right) \sqrt{L_y^2 + a^2} \arctan\frac{k\left(s\right)}{\sqrt{a^2+L_y^2}}}{6 a^2 L_y^3}.
	\end{aligned}
\end{equation}
%
Furthermore, this law allows us to find another special case, the interaction between a rectangle and a prism of infinite length, R-RP, by setting $L_y \rightarrow \infty$. This law, for general $m$, is already derived in the previous section using RIIPI and finding the limit of Eq.~\eqref{eq:RRPLy} serves as verification for our derivation approach. Indeed, the limit case of Eq.~\eqref{eq:RRPLy} for $L_y \rightarrow \infty$ returns the same expression as \eqref{RRPvdW}, which confirms that our derivations are consistent. 

Regarding the present singularity for $h_x \rightarrow h_y$, we have already addressed it in the previous subsection by finding the limit, see Eq.~\eqref{RRPvdWspec}. Let us now fix this singularity by appropriate transformations. In essence, we multiply the numerator and the denominator by carefully selected factors, see \hyperref[supsec]{Notebook 6}. The final result is singularity-free, but less elegant since it contains a correction term $T$:
\begin{equation}
	\begin{aligned}
		\Pi_{\operatorname{R-RP}}^6 &= \lim_{L_y\rightarrow\infty} \Pi_{\operatorname{R-RP_{L_y}}}^6  =Q\left(q_2\right)-Q\left(2 b_x+q_2\right)-Q\left(2b_y+q_2\right)+Q\left(2b_x+2b_y+q_2\right) + T,  
	\end{aligned}
	\label{eq:RRP6eleg}
\end{equation}
where
\begin{equation}
	\begin{aligned}
		W\left(a\right) &\coloneqq \sqrt{\left(h_x-h_y\right)^2+a^2} \\
		Q\left(a\right) &\coloneqq \frac{\pi}{12 a^2} \left[ \frac{\left(a^2+\left(h_x+h_y\right)^2\right)^{3/2}}{\left(h_x+h_y\right)^2}-W\left(a\right) \right]\\
		T &\coloneqq \frac{16 \pi b_x b_y^2 (b_x+b_y+q_2)}{3T_1 T_2 T_3} \\
		T_1 &\coloneqq (h_x-h_y)^2 + 2 b_y q_2 + q_2^2 +W(q_2) W(2b_y+q_2)\\
		T_2 &\coloneqq Wt(q_2)+W(2b_xq_2)-W(2b_y+q_2)-W(2b_x+2b_y+q_2)\\
		T_3&\coloneqq 4b_x(b_x+b_y+q_2)+W(q_2)W(2 b_y+q_2) + W(2b_x+q_2)W(2b_x+2b_y+q_2).
	\end{aligned}
\end{equation}
An additional test is to compare this fixed expression for R-RP vdW interaction with the one obtained in the previous subsection for $h_x \rightarrow h_y$. Indeed, if we insert $h_x=h_y$ in Eq.~\eqref{eq:RRP6eleg}, the expression \eqref{RRPvdWspec} follows, cf. \hyperref[supsec]{Notebook 6}. This further confirms the consistency of our derivations.

Next, it is straightforward to derive some additional special cases and compare them to those from Section \ref{secgeneral}. For example, if we let $h_y\rightarrow\infty$ in $\Pi_{\operatorname{R-RP}}^6$, a rectangle-(infinite) plate law follows. It is the same as the one in \eqref{eqPT-PT} for $m=6$, if we multiply it with  $2 h_x$, i.e.
\begin{equation}
	\label{eq44}
	\begin{aligned}
	\Pi_{\operatorname{R-PT}}^6 &= \lim_{h_y\rightarrow\infty} \Pi_{\operatorname{R-RP}}^6  =\frac{\pi h_x }{6}  \left(\frac{1}{q_2^2}-\frac{1}{\left(2 b_x + q_2\right)^2}-\frac{1}{\left(2 b_y + q_2\right)^2}+\frac{1}{\left(2 b_x +2 b_y + q_2\right)^2}\right). 
	\end{aligned}
\end{equation}
This comparison shows that the R-R law \eqref{eq:RR6} is sound and general since it returns well-known expressions as special cases. For example, it is now straightforward to obtain PT-PT, PT-HS, and HS-HS vdW laws from \eqref{eq44}.


\subsection{vdW interaction of two rectangles with common normal}

Throughout the paper, we were mainly concerned with the cases that can be used for the modeling of fiber interactions. To check our approach and extend it to different use cases, let us derive an expression for the interaction between two rectangles, as considered in \cite{1960derocco}. The authors in \cite{1960derocco} refer to these rectangles as \emph{sheets}. The main difference w.r.t.~our R-R law is the orientation -- these sheets share the same normal through the center and perpendicular to their planes. The integral to solve is similar as in previous subsections, but with two main differences: (i) the offset is $q_1=y_3-x_3$ and we must integrate over both coordinates, (ii) the gap $q_2=y_1-x_1$ is fixed and there is no integration along the $e_1$ axis, i.e.
\begin{equation}
	\begin{aligned}
		\Pi_{\operatorname{S-S}}^6&=\int_{-Lx}^{L_x} \int_{-L_y}^{L_y} \int_{-h_y}^{h_y}  \int_{-h_x}^{h_x} \left[q_2^2+\left(y_2-x_2\right)^2+\left(y_3-x_3\right)^2\right]^{-3}  \dd{x_2}\dd{y_2}\dd{y_3} \dd{x_3}. 
	\end{aligned}
	\label{recsheets1}
\end{equation}
The integration is not straightforward, and several simplifications of intermediate results are required in WM14, see \hyperref[supsec]{Notebook 6}. The final expression is
\begin{equation}
	\begin{aligned}
		\Pi_{\operatorname{S-S}}^6&= \frac{1}{2 q_2^4}\left[H\left(1,1\right)+H\left(1,-1\right)+H\left(-1,1\right)+H\left(-1,-1\right)\right], 
	\end{aligned}
	\label{recsheets2}
\end{equation}
where
\begin{equation}
	\begin{aligned}
		H(s,b) &\coloneqq sb \left(\frac{L(b) \left[2 k^2(s) +q_2^2\right]  \arctan\frac{L\left(b\right)}{\sqrt{k^2\left(s\right)+q_2^2}}}{\sqrt{k^2(s)+q_2^2}} + \frac{k(s) \left[2 L^2(b) +q_2^2\right]  \arctan\frac{k\left(s\right)}{\sqrt{L^2\left(b\right)+q_2^2}}}{\sqrt{L^2(b)+q_2^2}} \right) \\
		L(b)&\coloneqq L_x+bL_y.
	\end{aligned}
	\label{recsheets3}
\end{equation}
By letting $h_x \rightarrow h_y$ and $L_x \rightarrow L_y$ in this expression, we obtain the interaction between two rectangular sheets with the same dimensions $2h_y \times 2L_y$, i.e.
\begin{equation}
	\begin{aligned}
		\hat{\Pi}_{\operatorname{S-S}}^6&=\lim\limits_{L_x \rightarrow L_y} \lim\limits_{h_x \rightarrow h_y} \Pi_{\operatorname{S-S}}^6 = \frac{1}{ q_2^4} \left[\delta(h_y)+\delta(L_y)+\sigma(L_y,h_y)+\sigma(h_y,L_y)\right]
	\end{aligned}
	\label{recsheets4}
\end{equation}
where
\begin{equation}
	\begin{aligned}
		\delta\left(a\right) &\coloneqq -a q_2 \arctan \frac{2 a}{q_2} \\
		\sigma \left(a,b\right) &\coloneqq \frac{ a \left(8 b^2 +q_2^2\right)  \arctan\frac{2a}{\sqrt{4b^2+q_2^2}}}{\sqrt{4b^2+q_2^2}} 
	\end{aligned}
	\label{recsheets5}
\end{equation}
The obtained expression is the same as derived in \cite{1960derocco}, cf.~\hyperref[supsec]{Notebook 6}. Therefore, our expression for the interaction of two rectangular sheets with arbitrary dimensions \eqref{recsheets2} is general and accurate. Furthermore, if we let $L_y\rightarrow \infty$ and $h_y\rightarrow\infty$ in \eqref{recsheets2}, a plane figure-infinite plane law for $m=6$ follows, cf. Eq.~\eqref{eq1rule2} and \hyperref[supsec]{Notebook 1}.


\section{Application example}
\label{secapp}


As an application example of pre-integrated interaction potentials for numerical simulations, let us consider the case of peeling and pull-off between two elastic fibers that interact via the Lennard-Jones (LJ) potential. The example was originally analyzed using the approximate in-plane disk-disk law which does not provide satisfactory accuracy \cite{2021grill}. The results have been improved by employing an approximate disk-cylinder law \cite{2023grill} and ISSIP \cite{2024borkovićd}. Here, we compare the results obtained with ISSIP and $\operatorname{D-D_{app}}$, see Subsection \ref{secappDD}. 

\begin{figure}[h!]
	\centering
	\includegraphics[width=0.8\textwidth]{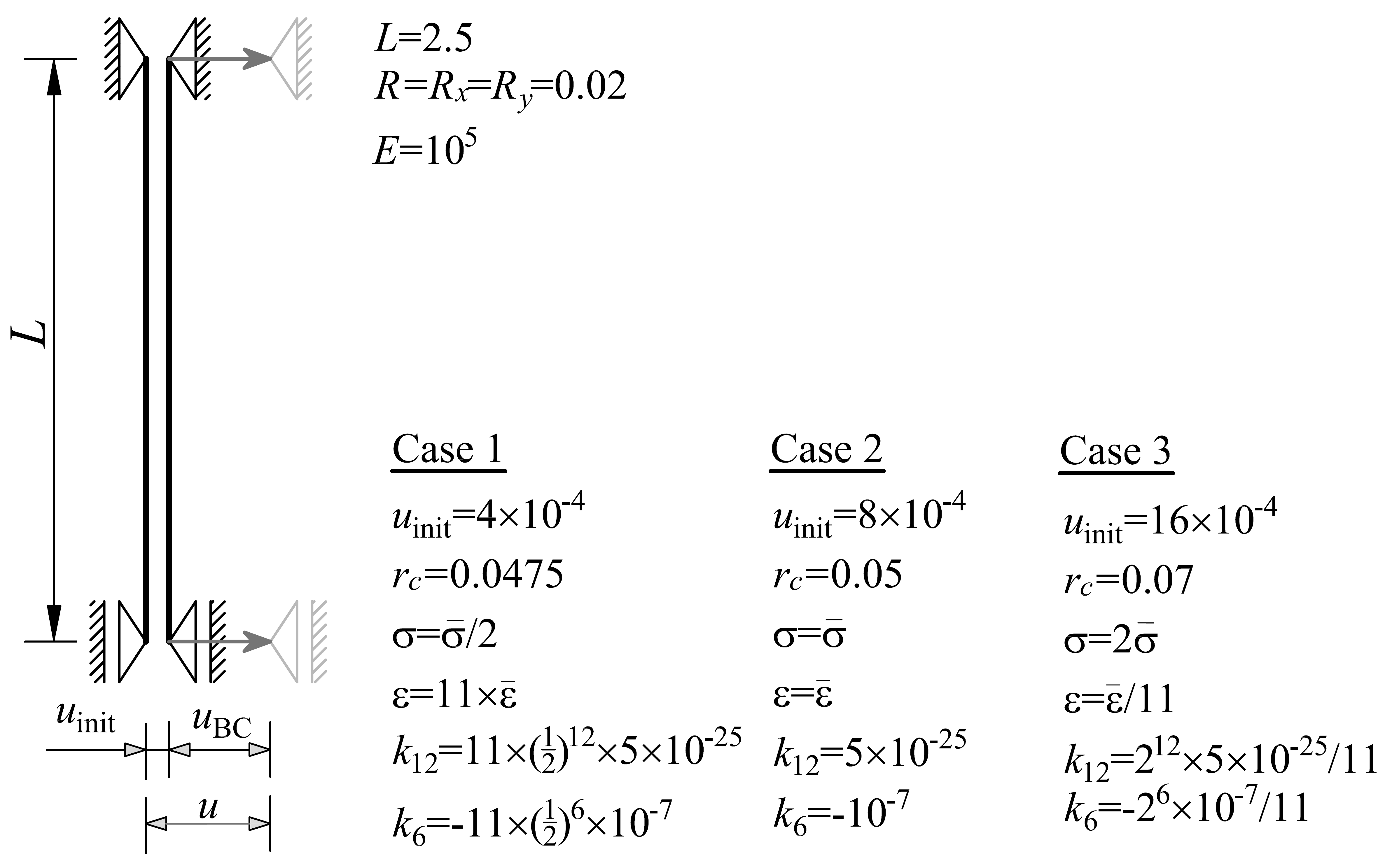}
	\caption{Setup and data used for the numerical experiments.}
	\label{fig:setup}
\end{figure}

The problem setup and input data are shown in Fig.~\ref{fig:setup}.
Two simply-supported elastic fibers with circular cross sections are initially set at a distance of $u_\text{init}$, which is slightly below the equilibrium distance.
A symmetric horizontal displacement of the supports, $u_\text{BC}$, is applied to the right fiber such that the distance between the supports of the interacting fibers is $u(t)=u_\text{init}+u_\text{BC}=t L$. Here, $t$ is a quasi-time parameter $t\in [0,1]$ and $L$ is the fiber's length, taken as $L=5$ in the original simulations of \cite{2023grill,2024borkovićd}. To reduce computational time without affecting the observed phenomenon, we set $L=2.5$ here. Due to applied non-homogeneous boundary condition $u_\text{BC}$, peeling between the fibers occurs. The peeling is followed by a pull between fibers and the simulation ends with a pull-off \cite{2024borkovićd, 2024borkoviće}. A straightforward solution approach is to integrate the P-P LJ potential over the volumes of both fibers. However, solving the two folded 3D integrals in such a highly nonlinear setting is computationally expensive. If we use a section-section (disk-disk) pre-integrated law, we need to integrate only two folded 1D integrals, which results in a substantial improvement in efficiency.

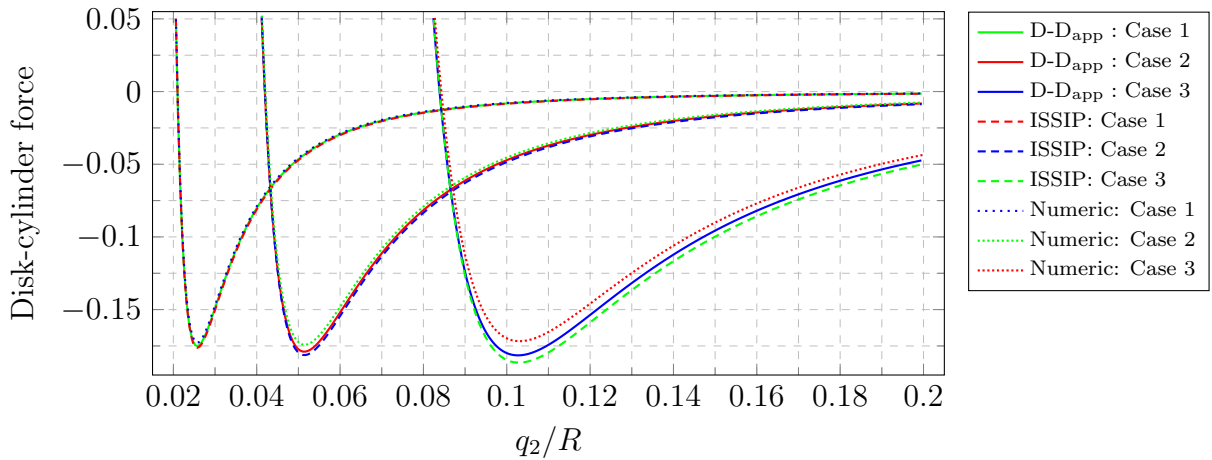
\begin{figure}[h!]
	\centering
	\begin{tikzpicture}
		\begin{axis}[
			xlabel = {$q_2/R$},
			ylabel = {Disk-cylinder force 
			},
			ylabel near ticks,
			legend pos=outer north east,
			legend cell align=left,
			legend style={font=\scriptsize},
			width=0.75\textwidth,
			height=0.40\textwidth,
			ymin = -0.195, ymax = 0.055,
			xmin=0.015,xmax=0.205,
			minor y tick num = 1,
			minor x tick num = 1,
			ytick distance=0.05,
			grid=both,
			xticklabel style={/pgf/number format/fixed, /pgf/number format/precision=2},
			yticklabel style={/pgf/number format/fixed, /pgf/number format/precision=2},
			clip=false]
			
			\addplot[green,thick] table [col sep=comma] {data/dataNumExampleLJF2DDappDist05x.csv};

			\addplot[red,thick] table [col sep=comma] {data/dataNumExampleLJF2DDappSigma1.csv};
			
			\addplot[blue,thick] table [col sep=comma] {data/dataNumExampleLJF2DappDistx2.csv};
			
			\addplot[red,densely dashed,thick] table [col sep=comma] {data/dataNumExampleLJF2ISSIPDist05x.csv};

			\addplot[blue,densely dashed,thick] table [col sep=comma] {data/dataNumExampleLJF2ISSIPSigma1.csv};
			
			\addplot[green,densely dashed,thick] table [col sep=comma] {data/dataNumExampleLJF2ISSIPDistx2.csv};
			
			\addplot[blue,thick,dotted] table [col sep=comma] {data/dataNumExampleLJF2DDExactDistx05.csv};
			
			\addplot[green,thick,densely dotted] table [col sep=comma] {data/dataNumExampleLJF2DDExactDistx1.csv};
			
			\addplot[red,thick,densely dotted] table [col sep=comma] {data/dataNumExampleLJF2DDExactDistx2.csv};
			
			\legend{$\operatorname{D-D_{app}}:$ Case 1,$\operatorname{D-D_{app}}:$ Case 2,$\operatorname{D-D_{app}}:$ Case 3,ISSIP: Case 1,ISSIP: Case 2,ISSIP: Case 3,Numeric: Case 1,Numeric: Case 2,Numeric: Case 3} 
		\end{axis}
	\end{tikzpicture}
	\caption{Disk-cylinder LJ force function for three considered cases with different physical constants. The approximate disk-cylinder force functions are obtained by analytically integrating two disk-disk laws: $\operatorname{D-D_{app}}$ and ISSIP. Highly accurate disk-cylinder force values are obtained by numerically integrating the exact P-C law.}
	\label{fig:LJpoten5}
\end{figure}

We consider three cases that differ in the values of the physical constants. The P-P LJ interaction potential can be represented as a function of two parameters: (i) the distance at which the potential is zero, $\sigma$, and (ii) the minimum value of the potential, $\epsilon$, i.e.
\begin{equation}
	\begin{aligned}
		\Pi_{\operatorname{P-P}}^{\operatorname{LJ}}=4\epsilon \left[\left(\frac{\sigma}{r}\right)^{12}-\left(\frac{\sigma}{r}\right)^6\right] = k_6 r^{-6} + k_{12} r^{-12}.
	\end{aligned}
	\label{eqLJex}
\end{equation}
The reference physical constants are adopted as in \cite{2024borkovićd}, $k_6=-10^{-7}$ and $k_{12}=5\times 10^{-25}$, which gives $\sigma= \bar \sigma =(200 \times 5^{5/6})^{-1}$ and $\epsilon= \bar \epsilon= -5 \times 10^9$. A numerical model with these parameters is designated as Case 2. To inspect the influence of equilibrium distance, we divide and multiply parameter $\bar \sigma$ by 2, and obtain Case 1 and Case 3, respectively. For the P-P LJ interaction potential, the change of $\sigma$ affects only the distance at which the potential is zero. However, for all interactions obtained by the integration of the P-P LJ potential, the change of $\sigma$ affects both the equilibrium distance and the maximum value of the attractive force. Although we apply disk-disk laws in our numerical simulations, it is more informative to consider disk-infinite cylinder interaction to scrutinize the effect of parameters $\sigma$ and $\epsilon$ on the interaction between fibers. Therefore, we integrate our $\operatorname{D-D_{app}}$ and ISSIP laws along the offset $q_1$ from $-\infty$ to $\infty$ and observe the disk-infinite cylinder laws. To focus on the influence of the equilibrium distance only, we multiply and divide parameter $\bar \epsilon$ by 11 for Case 1 and Case 3, respectively. As a result, the D-C equilibrium distances for the three consecutive cases differ by 2, while the maximum D-C attractive force is approximately the same. The D-C force functions are displayed in Fig.~\ref{fig:LJpoten5} for the adopted cases and both approximate disk-disk laws. 
%
Additionally, highly accurate values are calculated by numerical integration of the exact P-C laws defined in \eqref{eq:PC}. By inspecting these force plots, we notice that the differences between the numerical, $\operatorname{D-D_{app}}$, and ISSIP results increase with the equilibrium distance. This is expected behavior because the differences between the D-D, $\operatorname{D-D_{app}}$, and ISSIP laws increase with the gap, cf. Fig.~\ref{fig:DDAppvsISSIPcomp}.

The fibers are modeled as Bernoulli-Euler beams and the equilibrium equations are derived by the principle of virtual work \cite{2023borković}. Each fiber is spatially discretized into 80 quartic B-spline elements with $C^3$ interelement continuity. The equilibrium equations are solved by the Newton-Raphson method. Due to the large gradients of the interaction forces, 100 integration points per element are utilized. To improve efficiency, we employ a cutoff distance, $r_c$, in the present simulations. For Case 2, we have already concluded that the value $r_c=0.05$ provides good accuracy \cite{2024borkovićd}. Since the force range increases with the equilibrium distance, we have adopted the values $r_c=0.0475$ and $r_c=0.07$ for Case 1 and Case 3, respectively. 

\begin{figure}[h!]
	\centering
	\includegraphics[width=0.9\textwidth]{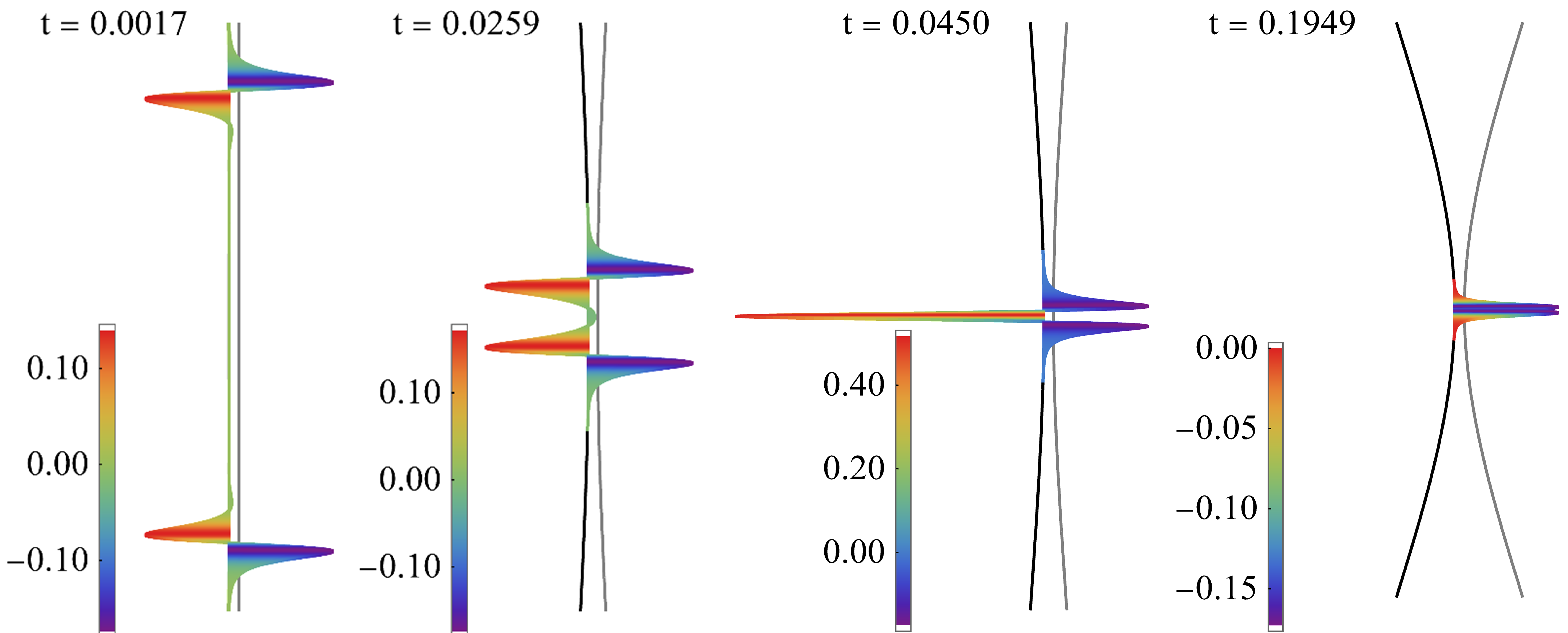}
	\caption{Normal component $f_2$ of the interaction force plotted on the left fiber for Case 1. Four characteristic configurations, calculated with the $\operatorname{D-D_{app}}$ law, are shown.}
	\label{fig:ExamNormDist05x}
\end{figure}

\begin{figure}[h!]
	\centering
	\includegraphics[width=0.9\textwidth]{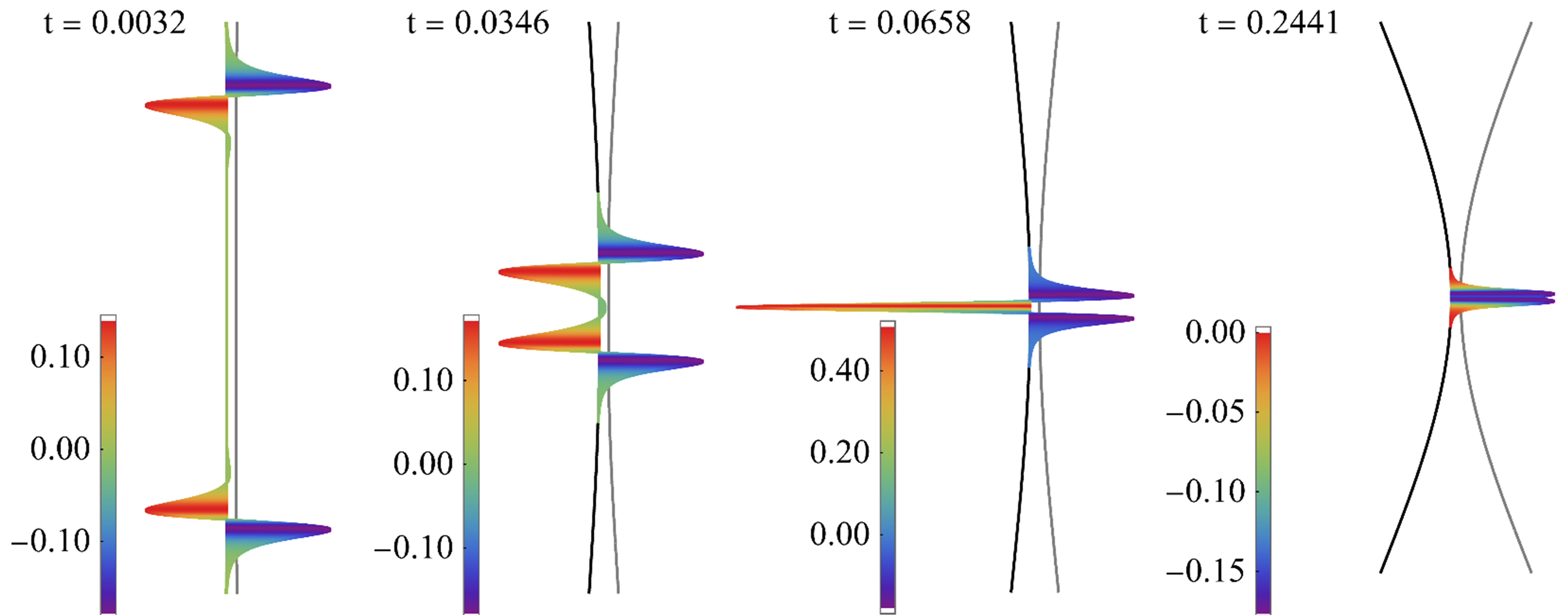}
	\caption{Normal component $f_2$ of the interaction force plotted on the left fiber for Case 2. Four characteristic configurations, calculated with the $\operatorname{D-D_{app}}$ law, are shown.}
	\label{fig:ExamNormDist1x}
\end{figure}

\begin{figure}[h!]
	\centering
	\includegraphics[width=0.9\textwidth]{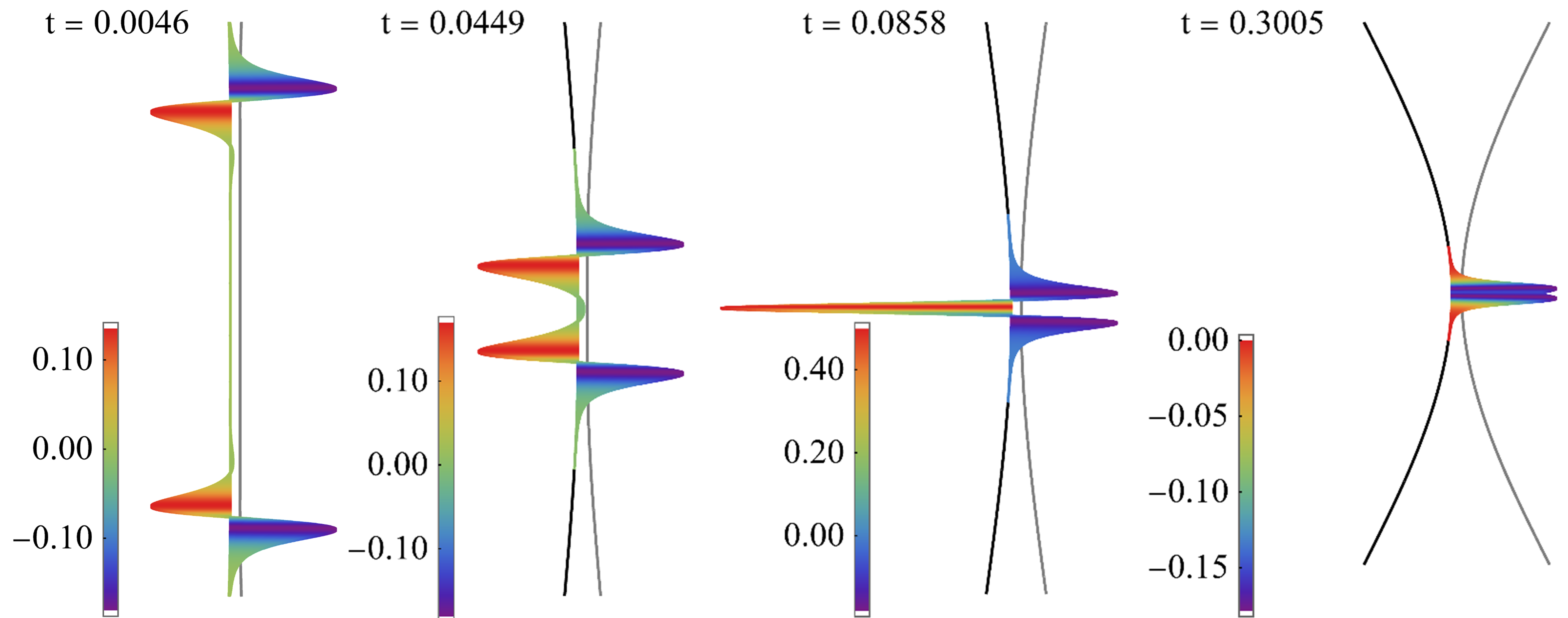}
	\caption{Normal component $f_2$ of the interaction force plotted on the left fiber for Case 3. Four characteristic configurations, calculated with the $\operatorname{D-D_{app}}$ law, are shown.}
	\label{fig:ExamNormDist2x}
\end{figure}

Let us consider the normal component of the distributed interaction force per unit length of a fiber, denoted $f_2$. For brevity, we refer to this quantity as the \emph{interaction force}. It is plotted on the left fiber for four characteristic configurations in Figs.~\ref{fig:ExamNormDist05x}, \ref{fig:ExamNormDist1x}, and \ref{fig:ExamNormDist2x}.
The peeling between fibers is represented in the first two plotted configurations. The third configuration represents the start of pulling, when the maximum repulsive force occurs. The fourth depicted configuration represents the instance before pull-of. For all three considered cases, the fibers behave similarly and the amplitudes of the distributed interaction force are close. However, significant differences exist in the net interacting forces due to the different force distribution lengths. An increase in the equilibrium distance of the LJ force increases the force range and the force distribution length, see also Fig.~\ref{fig:LJpoten5}. These different net interaction forces visibly affect the deformation of fibers. For example, Case 1 provides the smallest and Case 3 the largest bending of the fibers at pull-off.

\begin{figure}[h!]
	\centering
	\begin{tikzpicture}[spy using outlines=	{rectangle, magnification=2.0, connect spies}]
		\begin{axis}[
			xlabel = {$\xi$},
			ylabel = {Interaction force $f_2$},
			ylabel near ticks,
			legend pos=south east,
			legend cell align=left,
			legend style={font=\scriptsize},
			width=0.9\textwidth,
			height=0.5\textwidth,
			xmin = 0.44, xmax = 0.56,
			ymin = -0.1875, ymax = 0,
			minor y tick num = 1,
			minor x tick num = 2,
			xtick distance=0.02,
			xticklabel style={/pgf/number format/fixed, /pgf/number format/precision=2},
			yticklabel style={/pgf/number format/fixed, /pgf/number format/precision=2},
			clip=false,grid=both];
		
			\coordinate (spypoint) at (axis cs:0.5,-0.1662);
			
			\coordinate (magnifyglass) at (axis cs:0.462,-0.13);
			
			\addplot[green,thick] table [col sep=comma] {data/dataNumExampleF2DistributionDDappDist05x.csv};
			
		 	\addplot[red,thick] table [col sep=comma] {data/dataNumExDDappF2Distsig1.csv};
			
			\addplot[blue,thick,restrict x to domain=0.44:0.56] table [col sep=comma] {data/dataNumExampleF2DistributionDDappDist2x.csv};
			
			\addplot[orange,thick,densely dashed] table [col sep=comma] {data/dataNumExampleF2DistributionISSIPDist05x.csv};
			
			\addplot[black,thick,densely dashed] table [col sep=comma] {data/dataNumExISSIPF2Distsig1.csv};

			\addplot[purple,thick,densely dashed,restrict x to domain=0.44:0.56] table [col sep=comma] {data/dataNumExampleF2DistributionISSIPDist2x.csv};
			\legend{$\operatorname{D-D_{app}}:$ Case 1,$\operatorname{D-D_{app}}:$ Case 2,$\operatorname{D-D_{app}}:$ Case 3,ISSIP: Case 1,ISSIP: Case 2,ISSIP: Case 3} 
		\end{axis}
		\spy [lightgray, height=2.75cm,width=4.1cm] on (spypoint)
			in node[fill=white] at (magnifyglass);
	\end{tikzpicture}
	\caption{Distribution of the normal component of the interaction force $f_2$ at pull-off using the $\operatorname{D-D_{app}}$ and ISSIP laws for three considered cases.}
	\label{fig:LJpoten35}
\end{figure}
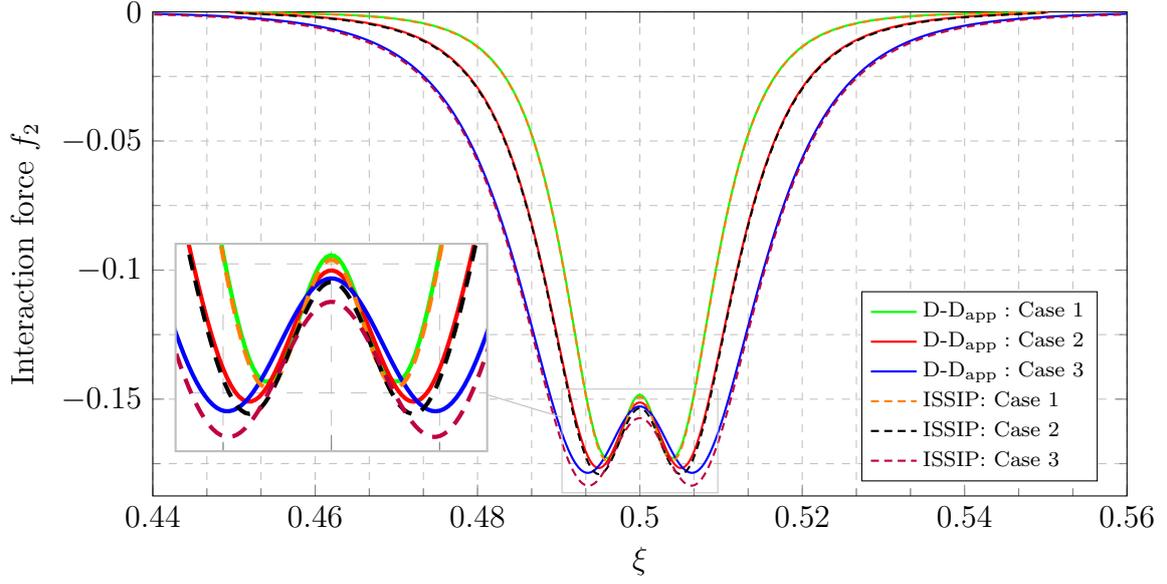

The distribution of the interaction force $f_2$ at pull-off is displayed in Fig.~\ref{fig:LJpoten35} as a function of the fiber's parametric length coordinate $\xi\in[0,1]$. 
The maximum values of the attraction force are in-line with the D-C interaction forces in Fig.~\ref{fig:LJpoten5}. As aforementioned, the force distribution length increases with the equilibrium distance. Regarding the two considered approximate disk-disk laws, the difference between them increases with the equilibrium distance, as predicted by the D-C force law in Fig.~\ref{fig:LJpoten5}.

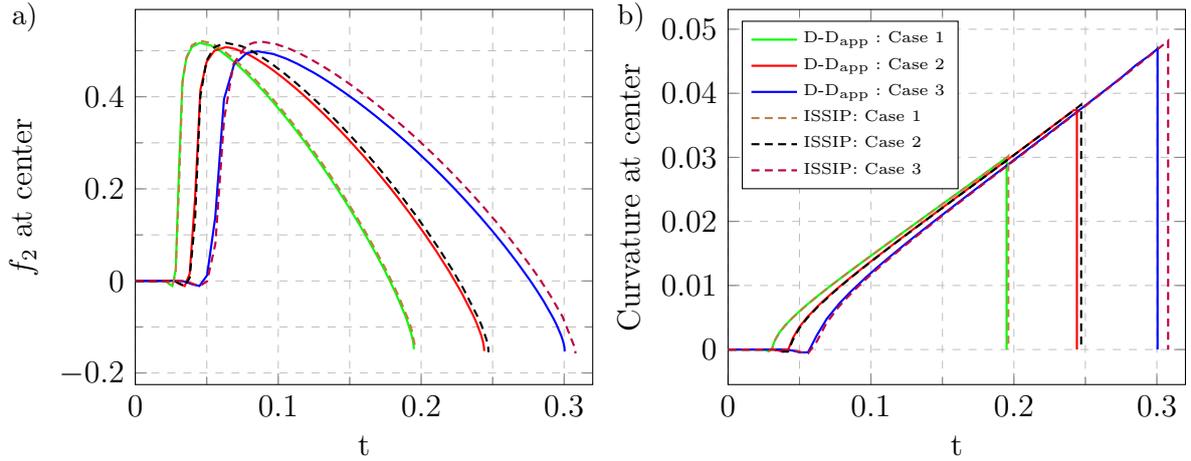
\begin{figure}[h!]
	\centering
	\begin{tikzpicture}
		\begin{axis}[
			xlabel = {t},
			ylabel = {$f_2$ at center},
			ylabel near ticks,
			legend pos=north east,
			legend cell align=left,
			legend style={font=\tiny},
			width=0.475\textwidth,
			xmin = 0, xmax = 0.32,
			minor y tick num = 1,
			minor x tick num = 1,
			xticklabel style={/pgf/number format/fixed, /pgf/number format/precision=2},
			yticklabel style={/pgf/number format/fixed, /pgf/number format/precision=2},
			clip=false,grid=both];
			\node [text width=1em,anchor=north west] at (rel axis cs: -0.3,1.1){\subcaption{\label{fig:a}}};
			
			\addplot[green,thick] table [col sep=comma] {data/dataNumExampleIntForceF2DDappDist05x.csv};
			
			\addplot[red,thick] table [col sep=comma] {data/dataNumExampleIntForceF2DDappDist1x.csv};
			
			\addplot[blue,thick] table [col sep=comma] {data/dataNumExDDappF2Dist2x.csv};
				
			\addplot[brown,thick,densely dashed] table [col sep=comma] {data/dataNumExampleIntForceF2ISSIPDist05x.csv};
				
			\addplot[black,thick,densely dashed] table [col sep=comma] {data/dataNumExampleIntForceF2ISSIPDist1x.csv};
		
			\addplot[purple,thick,densely dashed] table [col sep=comma] {data/dataNumExISSIPF2Dist2x.csv};
		\end{axis}
	\end{tikzpicture}
	\begin{tikzpicture}
		\begin{axis}[scaled ticks=false,
			xlabel = {t},
			ylabel = {Curvature at center},
			ylabel near ticks,
			legend pos=north west,
			legend cell align=left,
			legend style={font=\tiny},
			width=0.475\textwidth,
			xmin = 0, xmax = 0.32,
			minor x tick num = 1,
			xticklabel style={/pgf/number format/fixed, /pgf/number format/precision=3},
			yticklabel style={/pgf/number format/fixed, /pgf/number format/precision=3},
			clip=false,grid=both];
			\node [text width=1em,anchor=north west] at (rel axis cs: -0.27,1.1){\subcaption{\label{fig:b}}};
			
			\addplot[green,thick] table [col sep=comma] {data/dataNumExampleCurvatureDappDist05x.csv};
			
			\addplot[red,thick] table [col sep=comma] {data/dataNumExampleCurvatureDappDist1x.csv};
			
			\addplot[blue,thick] table [col sep=comma] {data/dataNumExampleCurvatureDappDist2x.csv};
			
			\addplot[brown,thick,densely dashed] table [col sep=comma] {data/dataNumExampleCurvatureISSIPDist05x.csv};
			
			\addplot[black,thick,densely dashed] table [col sep=comma] {data/dataNumExampleCurvatureISSIPDist1x.csv};
			
			\addplot[purple,thick,densely dashed] table [col sep=comma] {data/dataNumExampleCurvatureISSIPDist2x.csv};
			
			\legend{$\operatorname{D-D_{app}}:$ Case 1,$\operatorname{D-D_{app}}:$ Case 2,$\operatorname{D-D_{app}}:$ Case 3,ISSIP: Case 1,ISSIP: Case 2,ISSIP: Case 3} 
		\end{axis}
	\end{tikzpicture}
	\caption{Comparison of the $\operatorname{D-D_{app}}$ and ISSIP laws for Case 1, Case 2, and Case 3. a) The normal component of the interaction force, $f_2$, at center vs.~$t$. b) The curvature of the fiber's axis at center vs.~$t$. The $\operatorname{D-D_{app}}$ law is more accurate than the ISSIP law, see also Figs.~\ref{fig:DDAppvsISSIPcomp} and \ref{fig:LJpoten5}.}
	\label{fig:LJpoten4}
\end{figure}

The value of the interaction force and the curvature at the fiber's center, as a function of parameter $t$, are plotted in Fig.~\ref{fig:LJpoten4} for all three cases and both approximate disk-disk laws. The $\operatorname{D-D_{app}}$ law is more accurate than the ISSIP law, see also Figs.~\ref{fig:DDAppvsISSIPcomp} and \ref{fig:LJpoten5}.
In line with the previous considerations, significant differences between the cases are evident and the differences between the results obtained with the approximate disk-disk laws increase with the equilibrium distance of the LJ force, from Case 1 to Case 3.


Regarding the efficiency, the computational time for both disk-disk laws is practically the same, since it is governed by the evaluation of a hypergeometric function. 
Therefore, for efficient numerical simulations that require increased accuracy, the new approximate $\operatorname{D-D_{app}}$ law is preferable over the ISSIP. This is especially important for cases with relatively large equilibrium distances since the difference between the laws increases with the gap. 

Regarding the accuracy, the comparison of the employed approximate laws with the exact LJ D-C force in Fig.~\ref{fig:LJpoten5} shows that the error of the $\operatorname{D-D_{app}}$ law also increases with the equilibrium distance. Therefore, a derivation and an implementation of a more accurate disk-disk law is desirable for simulations involving large gaps. The exact D-D law, derived in Subsection \ref{secexactDD}, is a good starting point, but the mentioned numerical issues should be addressed. Furthermore, an appropriate steric disk-disk potential ($m=12$) should be derived.

\section{Conclusions}

In general, interaction potentials modeled as inverse-power laws of the point-pair distance cannot be integrated analytically over arbitrarily shaped interacting bodies. Even for simple geometries, constant densities, and rigid bodies, the problem is very complex and general analytical solutions do not exist. This problem is tackled here for several types of geometries and arbitrary interaction exponent $m$ but with a focus on vdW attraction, i.e. $m=6$. The geometries are chosen such that the results can be used as pre-integrated expressions for numerical simulations of interactions between deformable bodies that resemble fibers. We used \emph{Wolfram Mathematica} 14 for the pre-integration.

We have thus obtained the following potentials, which, to the best of our knowledge, are new: 
\begin{itemize}
\item A representation of the in-plane disk-disk interaction in the form of a generalized hypergeometric function for the case of disks with equal radii.
\item The exact disk-disk law for $m=6$ and a new approximate disk-disk law.
\item An approximate point-cylinder law for arbitrary even $m$.
\item An approximate disk-cylinder law for $m=6$.
\item The exact disk-plate law for arbitrary exponent.
\item The exact rectangle-rectangle in-plane law for arbitrary $m\ge4$.
\item The exact rectangle-rectangle law for $m=6$.
\item The exact sheet-sheet law with arbitrary dimensions for $m=6$.
\end{itemize}

Future research should deal with the arbitrary orientation of sections and the application of the pre-integrated potentials in numerical simulations.

\section*{Supplementary data}
\label{supsec}

The supplementary Wolfram Mathematica notebooks are curated in the online dataset with doi: \href{https://doi.org/10.3217/ytdcx-kyx08}{10.3217/ytdcx-kyx08}.

\section*{Acknowledgments}


This research was funded in part by the Austrian Science Fund (FWF) \href{https://doi.org/10.55776/P36019}{10.55776/P36019}. For the purpose of open access, the authors have applied a CC BY public copyright licence to any Author Accepted Manuscript version arising from this submission.

\section*{Appendix}

\appendix

\section{Infinite series approach for in-plane disk-disk interaction}

\label{appendix}
\setcounter{equation}{0}
\renewcommand\theequation{A\arabic{equation}}

A step-by-step derivation of the in-plane disk-disk interaction law in the form of an infinite series is presented, see \cite{1972langbein}. The relative polar coordinate system (RPCS2) of Fig.~\ref{fig:intro1}c is utilized and
%
%
the integral to solve is
\begin{equation}
	\label{eqapp1}
\begin{aligned}
\bar{\Pi}_{\operatorname{D-D_{IP}}}^m&=\int _{A_x}\int _{A_y}\frac{1}{p^m}\dd{A_y}\dd{A_x} = 4 \int _{0}^{R_x}\int _{0}^\pi \int _{0}^{R_y}\int _{0}^\pi \frac{1}{p^m} r_x r_y \dd{\varphi_y}\dd{r_y} \dd{\varphi_x}\dd{r_x}.
\end{aligned}
\end{equation}
%
The distances $p$ and $t$ follow from the cosine theorem, i.e.
\begin{equation}
	\begin{aligned}
		p^2=t^2+r_y^2-2 t r_y \cos \varphi_y \\
		t^2 = d^2 + r_x^2 - 2 d r_x \cos \varphi_x.
	\end{aligned}
\end{equation}
With this parameterization, one part of the integral \eqref{eqapp1} reduces to the following table integral, see \cite{2014zwillinger} (p.~409, n.~3.665.2):
\begin{equation}
		\begin{aligned}
			\int_{0}^{\pi} \frac{\sin^{2\mu-1}x}{\left(1+2a\cos x + a^2\right)^m} \dd{x} &= B \left(\mu,\frac{1}{2}\right) \,_2F_1 \left(m,m-\mu+\frac{1}{2}; \mu+\frac{1}{2};a^2\right), \\
			&\text{with} \; \text{Re} \mu >0 \land \abs{a}<1,
		\end{aligned}
		\label{eq:app2}
\end{equation}
where $B(w,v)=\frac{\Gamma (w) \Gamma (v)}{\Gamma (w+v)}$ is the beta function. 
In our case $\mu=1/2$ and the table integral \eqref{eq:app2} simplifies to
\begin{equation}
	\begin{aligned}
		\int_{0}^{\pi} \frac{1}{\left(1+2a\cos x + a^2\right)^m} \dd{x} &= B \left(\frac{1}{2},\frac{1}{2}\right) \,_2F_1 \left(m,m; 1;a^2\right) = \pi \,_2F_1 \left(m,m; 1;a^2\right).
	\end{aligned}
	\label{eq:app4}
\end{equation}

Let us first integrate \eqref{eqapp1} w.r.t.~the area of disc $X$, i.e. consider the P-D$_{\text{IP}}$ interaction
\begin{equation}
	\begin{aligned}
\bar{\Pi}_{\operatorname{P-D_{IP}}}^m&= 2 \int_{0}^{R_x} \int_{0}^{\pi} \frac{r_x}{p^m} \dd{\varphi_x} \dd{r_x} = 2 \int_{0}^{R_x} r_x \int_{0}^{\pi} (t^2+r_x^2-2 t r_x \cos \varphi_x)^{-\frac{m}{2}}  \dd{\varphi_x} \dd{r_x}.
	\end{aligned}
\end{equation}
By introducing the substitution $a=-\frac{r_x}{t}$, the integral becomes
\begin{equation}
	\begin{aligned}
		\bar{\Pi}_{\operatorname{P-D_{IP}}}^m&= 2 \int_{0}^{R_x} r_x \int_{0}^{\pi} \left[ t^2 \left(1+a^2+2 a \cos \varphi_x\right)\right]^{-\frac{m}{2}}  \dd{\varphi_x} \dd{r_x} \\
		&= 2 t^{-m} \int_{0}^{R_x} r_x  \int_{0}^{\pi} \left(1+a^2+2 a \cos \varphi_x\right)^{-\frac{m}{2}}  \dd{\varphi_x} \dd{r_x},
	\end{aligned}
\end{equation}
and we can apply the table integral \eqref{eq:app4}, i.e.
\begin{equation}
	\begin{aligned}
		\bar{\Pi}_{\operatorname{P-D_{IP}}}^m&=  2 \pi t^{-m} \int_{0}^{R_x} r_x  \, _2F_1\left(\frac{m}{2},\frac{m}{2};1;\frac{r_x^2}{t^2}\right) \dd{r_x} = 2 \pi t^{-m} \int_{0}^{R_x}   \sum_{n=0}^{\infty} \frac{\left(\frac{m}{2}\right)_n \left(\frac{m}{2}\right)_n}{\left(1\right)_n n!} \frac{r_x^{2n+1}}{t^{2n}} \dd{r_x} \\
		&= 
		\pi    \sum_{n=0}^{\infty} \frac{\left(\frac{m}{2}\right)_n \left(\frac{m}{2}\right)_n}{n! \left(n+1\right)!} \frac{R_x^{2n+2}}{t^{2n+m}} = \pi R_x^2 \, t^{-m}\, _2F_1 \left(\frac{m}{2},\frac{m}{2};2;\frac{R_x^2}{t^2}\right) \; \text{for} \; m>2.
	\end{aligned}
	\label{eq:appPDIP}
\end{equation}
Equation \eqref{eq:appPDIP} represents the P-D$_{\text{IP}}$ law for general $m>2$ in the form of a hypergeometric function.  

Now $\bar{\Pi}_{\operatorname{P-D_{IP}}}^m$ is left to be integrated w.r.t.~the area of disc $Y$, i.e.
\begin{equation}
	\begin{aligned}
		\bar{\Pi}_{\operatorname{D-D_{IP}}}^m&=\int_{A_y} \bar{\Pi}_{\operatorname{P-D_{IP}}}^m \dd{A_y} = 2 	\pi    \sum_{n=0}^{\infty} \frac{\left(\frac{m}{2}\right)_n \left(\frac{m}{2}\right)_n}{n! \left(n+1\right)!} R_x^{2n+2} \int_{0}^{R_y} \int_{0}^{\pi}  \frac{1}{t^{2n+m}} r_y \dd{\varphi_y} \dd{r_y}
	\end{aligned}
\end{equation}
%
%
and the procedure is analogous to the P-D$_{\text{IP}}$ integral, i.e.
\begin{equation}
	\begin{aligned}
		\bar{\Pi}_{\operatorname{D-D_{IP}}}^m&=2 \pi^2 \sum_{n=0}^{\infty} \frac{\left(\frac{m}{2}\right)_n \left(\frac{m}{2}\right)_n}{n! \left(n+1\right)!} \frac{R_x^{2n+2}}{d^{\left(2n+m\right)}}  \int_{0}^{R_y} r_y  \, _2F_1 \left(n+\frac{m}{2},n+\frac{m}{2};1;\frac{r_y^2}{d^2}\right) \dd{r_y} \\
		&= 2 \pi^2 \sum_{n=0}^{\infty} \frac{\left(\frac{m}{2}\right)_n \left(\frac{m}{2}\right)_n}{n! \left(n+1\right)!} R_x^{2n+2}  \int_{0}^{R_y} \sum_{k=0}^{\infty} \frac{\left(n+\frac{m}{2}\right)_k \left(n+\frac{m}{2}\right)_k}{\left(1\right)_k k!} \frac{r_y^{2k+1}}{d^{2k+2n+m}} \dd{r_y} \\
		&= \pi^2 \sum_{n,k=0}^{\infty} \frac{\left(\frac{m}{2}\right)_n \left(\frac{m}{2}\right)_n \left(n+\frac{m}{2}\right)_k \left(n+\frac{m}{2}\right)_k}{n! \left(n+1\right)! k! \left(k+1\right)!} \frac{R_y^{2k+2}R_x^{2n+2}}{d^{2k+2n+m}}\; \text{for} \; m>\frac{7}{2}.
	\end{aligned}
\end{equation}
%
%
%
If we substitute Pochhammer symbols with Gamma functions and introduce reduced radii $R_x/d$ and $R_y/d$, the expression becomes
\begin{equation}
	\begin{aligned}
		\bar{\Pi}_{\operatorname{D-D_{IP}}}^{m}= \frac{\pi^2}{d^{m-4}\Gamma^2\left(\frac{m}{2}\right)}    \sum_{n,k=0}^{\infty} \frac{\Gamma^2 \left(k+n+\frac{m}{2}\right)}{k! n! \Gamma \left(k+2\right) \Gamma \left(n+2\right)} \left(\frac{R_x}{d}\right)^{2k+2} \left(\frac{R_y}{d}\right)^{2n+2}.
	\end{aligned}
\end{equation}
This result is the same as in \cite{1972langbein} with one difference: the authors in \cite{1972langbein} have eliminated the first terms of both series. This expression can also be represented via $_2F_1$ function, i.e.
\begin{equation}
	\begin{aligned}
		\bar{\Pi}_{\operatorname{D-D_{IP}}}^{m} &= \frac{\pi^2 R_x^2 R_y^2}{d^{m}\Gamma^2\left(\frac{m}{2}\right)}    \sum_{n=0}^{\infty} \frac{\Gamma^2 \left(n+\frac{m}{2}\right)}{\Gamma \left(n+1\right) \Gamma \left(n+2\right)} \left(\frac{R_y}{d}\right)^{2n} \, _2F_1 \left(n+\frac{m}{2},n+\frac{m}{2},2,\frac{R_x^2}{d^2}\right).
	\end{aligned}
	\label{appDDIP}
\end{equation}
Although the $_2F_1$ function is, in essence, an infinite series, this representation can significantly improve the computational time since WM14 calculates $_2F_1$ function in a very efficient way.

\printbibliography


\end{document}